\documentclass[12pt,preprint]{aastex}
\def\smas{$M_\odot$}

\def\2foe{$2\times10^{51}$ ergs}
\def\3foe{$3\times10^{51}$ ergs}

\begin{document}
\bibliographystyle{apj}

\title{ A Parameter Study of Type II Supernova Light Curves Using \\6 M${_\odot}$ He Cores}
\author{Timothy R. Young}
\affil{Physics Department, University of North Dakota}
\begin{abstract}
Results of numerical calculations of Type II supernova light curves
are presented.  The model progenitor stars have 
6 $M{_\odot}$ cores
and various envelopes, originating from a numerically evolved 20 $M{_\odot}$ star.  Five parameters that affect the light curves
are examined: the ejected mass, the progenitor radius, the explosion
energy, the $^{56}$Ni mass, and the extent of $^{56}$Ni mixing. The
following affects have been found: 1) the larger the progenitor radius
the brighter the early--time light curve, with little affect on the
late--time light curve, 2) the larger the envelope mass the fainter
the early light curve and the flatter the slope of the late light
curve, 3) the larger the explosion energy the brighter the early light
curve and the steeper the slope of the late light curve, 4) the larger
the $^{56}$Ni mass the brighter the overall light curve after 20 to 50
days, with no affect on the early light curve, 5) the more extensive
the $^{56}$Ni mixing the brighter the early light curve and the
steeper the late light curve. The primary parameters affecting 
the light curve shape are the progenitor radius and 
the ejected mass. The secondary parameters are the explosion energy, 
$^{56}$Ni mass and $^{56}$Ni mixing. I find that while in principle 
the general shape and absolute magnitude of a light curve indicate 
a unique set of parameters, in practice it is difficult to avoid 
some ambiguity in the parameters.  I find that the nickel--powered 
diffusion wave and the recombination of helium produce a prominent 
secondary peak in all our calculations. The feature is less prominent 
when compositional mixing, both $^{56}$Ni mixing and mixing between the 
hydrogen and helium layers, occurs. The model photospheric temperatures 
and velocities are presented, for comparison to observation.
\end{abstract}

\section{Introduction}
Type II supernovae (SNe) are classified by the presence of H in their spectra and sub-classes are based on their observed light curves. They are most likely caused from the gravitational collapse of a main sequence stars with M $>$ 8-10 $M_\odot$.
The differences in Type II light curves was 
first pointed out by Pskovskii (1978) by introducing the 
beta parameter which resulted in a continuous classification based on the slope of the light 
curve. \nocite{p78} Another study showed that light curves 
can be differentiated into two distinctive sub-classes, plateaus and linears \citep{bcr79}. Pata et al. (1994),  \nocite{pbct94} using a
 multivariable factor analysis, introduced a new 
classification based on an absolute peak magnitude vs. $\beta_{100}$ graph resulting in three 
sub-classes; Bright (includes both plateaus and linears), Normal (includes both
plateaus and linears), and Faint.\

Good observations of Plateau SNe such as SN 1969L, SN 1987A, SN 1993J and recently SN 1999em and SN 1999gi are usually followed by extensive theoretical light curve (LC) studies. After SN 1969L LC 
calculations correctly predicted features 
found in the observations \citep{gn69,gn76,gin71}. Later more refined analytic treatments and 
hydrodynamical codes were developed \citep{fa77,ck79}
and applied to red supergiant progenitors.
\cite{ln83} conducted a numerical parameter study to
 examine how the ejected mass, progenitor radius, and explosion energy affect the plateau duration and magnitude of the light curve and photospheric velocity of material.
They showed that relations
between explosion energy, envelope mass, and progenitor radius can be 
obtained from observations of plateau duration, 
the absolute magnitude and the photospheric velocity. We are currently preparing a paper investigating a comparison between supernovae of evolved numerical stars and polytrope stars using the method \citep{yj04}.\

More recently Hamuy (2003) \nocite{h03} has conducted an observational parameter study based on properties of 24 type II supernova spectra and light curves. The author finds correlations between Ni mass and plateau luminosity, large ranges in all parameters, and a continuum in the parameter space for type II plateaus. Other studies of type II plateaus have found similar results. One study presents a technique of determining Ni mass based on H$\alpha$ luminosity at nebular phases \citep{ecd04}. As mentioned above we will be publishing a paper comparing the light curves of numerical model explosions to those presented in \cite{ln83,ln85} and compare the least-squares fit formulas in their paper to those found in this study \citep{yj04}. 

Many papers on SN 1987A have been published on LC modeling and references 
can be found in review articles  \citep{abkw89,in89}. 
In order to obtain a good fit to SN 1987A variations in envelope masses, explosion energies, 
progenitor radii, and mixing of both hydrogen-helium and $^{56}$Ni were explored but only in 
a limited parameter space \citep{w88,nsky91,sn90,u93}. Podsiadlowski et al. (1992) in an attempt to find variations in Type II progenitors 
examined binary systems and showed it is possible to find many different scenarios. 
Hsu et al. (1993) \nocite{hjrp94}, using hydrodynamical models, examined Type II light curves 
by varying the envelope mass and explosion energy. 

The unusual LC of 1993J produced interest in possibly new progenitors. Many 
studies of SN 1993J conclude that the progenitor was a 4 $M_\odot$ helium core with an 
envelope of about 0.2 $M_\odot$ of hydrogen, a radius of about 300 R$_\odot$ and 
belonged to a binary system 
\citep{ybb95,ssknsy94,nsskys93,phjr93,bbpt94}. A more extensive study on SN 1993J was conducted by
Blinnikov et al. (1998) \nocite{beb98} comparing calculations from two different numerical codes by analyzing the evolution and explosion of the progenitor. Young
et al.(1995), using the cepheid distance to M81 \nocite{f94} (Freedman et al., 1994), found a lower limit
to the Ni mass of 0.1 \smas. Using the X-ray light curve \nocite{sn95} Suzuki and Nomoto (1995) found a upper limit
to explosion energy to be 1$\times10^{51}$ ergs. Recently the massive companion star for SN 1993J was observed 10 years after the explosion, confirming the suggested binary system with two similar mass stars \citep{msk04}. Superior observations of supernova progenitors are now starting to constraining model progenitors and making light curve analysis more precise, e.g. SN 2001du \cite{smgt03}.

Two well observed SNe 1999em and 1999gi have been extensively studied. The nature of the progenitor of SN1999em is constrained to be around 12 \smas \citep{sgt2001}. The upper mass limit of SN 1999gi has been revised to 15 \smas. Both these progenitor masses are similar to that found in numerical fits to the light curves of SN 1999em and SN 1999gi \citep{y03,ytj04}. 
As observations become more complete type II plateaus have a potential to be used as distance indicators. The distances to both SNe were found by using the expanding photosphere method (EPM) \citep{l01,l02}. Other studies examined type IIs as possible distance indicators using expansion velocities that correlate with the bolometric luminosity \citep{hp02}.

 Young and 
Branch (1989) \nocite{yb89} compared observed Type IIp light curves on an absolute magnitude scale 
and found a large spread in absolute magnitude, their plateau duration, and slope of the 
tails, attributing it to differences
in the physical properties of the progenitor. These studies indicate that the parameter space is quite large for the progenitor 
of Type II SNe and it is difficult to say what type of progenitor will explode or not. 
What is needed is a parameter study that 
encompasses all variables and all combinations, but this is unfeasible. Therefore the 
aim of this and future studies
is to have a standard SN model with parameters that are consistent with knowledge of progenitor
stars, explosion mechanisms, and hydrodynamics. It is then of interest to see if all 
observed light curves can fit into this model and how far in parameter space they deviate 
from the standard model. In this study the standard model is a 20 \smas \ main sequence star
that has gone through a wind mass loss of 2 \smas, left a neutron star of 1.4 \smas, thus 
ejecting about 16 \smas \ with a progenitor radius of 3$\times10^{13}$ cm. The energetics of the simulation has a total energy of 1$\times10^{51}$ ergs  or 1 foe (ten to the fifty one ergs) and ejects 0.07 \smas \ $^{56}$Ni mixed throughout the 6 \smas \ He core. 
The standard model is similar to SN 1987A except for the larger radius and less extensive mixing.
In this paper the parameter space around this model is explored.\\

In this paper I show the results of how varying each parameter can influence the shape 
and absolute magnitude of Type II light curves out to 400 days. The five parameters 
explored in this study are; the 
progenitor radius, envelope mass, explosion energy, $^{56}$Ni mass, and $^{56}$Ni mixing.
All light curves calculated in this paper use the numerically 
evolved 20 $M{_\odot}$ \ main sequence model with a 6 \smas \ 
helium core from  Woosley and Weaver (1980).
The envelope parameters, mass and radius, are varied using homology transformations (section 2).
The models are then exploded in a one dimensional, flux-limited hydrodynamical code
with a simple prescription for gamma-ray deposition (section 3). The bolometric light curves are
calculated and plotted on an absolute magnitude scale (section 4). The results are discussed in section 5.\

\section{Initial Models}
For the initial models of all explosions I use the 6 $M{_\odot}$ helium core from 
Woosley and Weaver (1980) \nocite{ww80}, originally a 20 $M{_\odot}$ main sequence star. The
 envelope mass and radius are subsequently modified in a systematic way. A total of 8 models
were constructed that are identified by a specific mass and radius (Table I).
In this study three different envelope masses were used producing total masses of
8, 12, and 16 $M{_\odot}$. For each envelope mass, three different radii where used 43 R$_\odot$ ($3\times10^{12}$ cm)
, 430 R$_\odot$ ($3\times10^{13}$ cm), and 4300 R$_\odot$ ($3\times10^{14}$ cm). The original H rich envelope 
was modified by homologous transformations to give the various masses and radii \citep{s58,c39}. 

For a homologous transformation in radius: 
$$ R' = x R.$$
 where R is the old radius, R$'$ is the new radius, 
and $x$ is the percent changed.
For a homologous transformation in mass : 
$$\frac{1-\beta_1}{\mu_1^4\beta_1^4} = \frac{1-\beta_o}{\mu_o^4\beta_o^4}\left(
\frac{M_1}{M_o}\right)^2$$
where $\beta_o$ and $\beta_1$ are the ratio of the gas pressure to the total pressure for the 
star before and after the homologous transformation. $M_o$ and $M_1$ are initial and final mass,
and $\mu_o$ and $\mu_1$ are the mean molecular weights for the initial and final 
configuration. Following either of these calculations the remaining physical variables, $\rho$ and T, 
must be solved to ensure hydrostatic equilibrium. The homologous transformation is
performed on the hydrogen envelope only and then matched to the He core.
The core and various envelopes were then re-zoned to ensure a continuous density, temperature and 
radius. Eight progenitor models in all were constructed for this study each 
having 170 zones in order to have consistent results for opacity floors (opacity minimum) and 
gamma-ray deposition which are used to calculate zone dependent quantities like the photosphere and 
the gamma-ray contribution to the luminosity (see section 3). The innermost 1.4 $M_\odot$ is assumed 
to form a neutron star and is removed from the core, but set as an inner boundary condition
in the explosions.\\
The constructed models are in hydrostatic equilibrium but not necessarily in radiative equilibrium. This 
should not be a problem since the shock moves through the star in 6 to 240 hours. Radiative equilibrium is important concerning 
the validity of the model stars representing real stars, but the point of this study is to 
compare various simple models and see how the LC responds. \\
Three different final explosion energies, $1\times10^{51}$, $2\times10^{51}$, and $3\times10^{51}$ ergs
are obtained by artificially placing the required energy, divided between kinetic and thermal, in the 
first few mass zones. This procedure is justifiable since the explosion mechanism for Type II SNe 
is beyond the scope of this paper. Furthermore, the type of shock is only 
important in the nucleosynthesis (Aufderheide 1991) which is not included in our study.
 
\section{Numerical Method and Gamma-Ray Deposition}
The radiation-hydrodyanmics code given to the author \citep{w92,sw84} contains the following attributes;
 spherically symmetric, 1-D, Lagrangian, flux-limited code which uses pseudo-viscosity 
and determines opacities calculated for hydrogen, helium, and oxygen rich layers. Modifications included a geometric dilution model for 
gamma-ray deposition, and updated opacity tables \citep{y94}. Rayleigh-Taylor mixing
and nuclear reactions are not included in the calculations. The Rosseland mean opacities are 
tabulated as functions of composition, temperature, and density,assuming local thermodynamic equilibrium.
Two effects not included in the Rosseland mean are line opacities and non-thermal excitation or 
ionization from gamma rays. In order to account for these affects an opacity minimum (opacity floor) is set at 
0.25 cm$^2$ g$^{-1}$ for the helium rich core and 0.01 cm$^2$ g$^{-1}$ for the hydrogen rich 
envelope.\\
The LC calculation is done in two steps, first the hydrodynamics is calculated, 
then in a second step the gamma-ray deposition is included in the computation. 
 The prescription for absorbed
gamma rays requires calculating a deposition function following Sutherland and 
Wheeler (1984) \nocite{sw84} at interval times of 3.5$\times10^6$ sec from 0 to 400 days. 
The deposition function calculates the diffusion of gamma-rays from the location they are produced 
to regions outside the Ni distribution, including gamma-rays escaping the ejecta.
In the first step one obtains the time evolution of density and volume of the ejecta that is necessary to calculate 
the absorbed gamma rays.

The energy produced 
by the gamma rays is not enough to affect the hydrodynamics and thus is not necessary to be included in the primary run. This saves computation time, avoiding a calculation of the deposition function at each time step.\\
The $^{56}$Ni mass is varied using 3 different masses of 0.035, 0.07, and 0.14 $M{_\odot}$ which are
mixed to three different regions; less than 0.3 $M{_\odot}$, 6 $M{_\odot}$ (He core), and 11 $M{_\odot}$ (core plus half the 
envelope, 7 $M{_\odot}$ for model G). The Ni distribution is a step function resulting in 
a uniform spherical distribution of $^{56}$Ni. As the ejecta becomes thin, the 
gamma rays deposit energy
in a symmetric sphere outside the $^{56}$Ni distribution. The mixing of $^{56}$Ni is artificial since we
are not mixing the composition along with it, but the goal of this study is not to create a rigorous model, only
to explore the major parameters that influence the LC. However it has been shown that the mixing of
H into the He layer does affect the LC as seen in SN1987A \citep{abkw89,in89}. The energy produced by the decay of $^{56}$Ni and $^{56}$Co is
\begin{equation}
S(t_{days})= \frac{3.4\times10^{11}e^{-t/\tau_{Ni}}}{\tau_{Ni}}+\frac{7.5\times10^{11}}
{\tau_{Co}-\tau_{Ni}}(e^{-t/\tau_{Co}}-e^{-t/\tau_{Ni}})\ ergs\ g^{-1}s^{-1}
\end{equation}
where S is the amount of energy per gram per second produced from the Ni-Co-Fe decay, t is time in days, 
$\tau_{Ni}$=8.8 days and $\tau_{Co}$=113.6 days and the amount being deposited in the SN ejecta ($S_{dep}$) is 
\begin{equation}
S_{dep}(t_{days}) = Dep \times S(t_{days})\ ergs\ g^{-1}s^{-1}
\end{equation}
where Dep is the deposition function.
In it is useful to introduce a gamma-ray ``photosphere'' (hereafter gammasphere), 
similar to how the actual photosphere is defined. This is taken to 
be $\tau_{\gamma}$ = 2/3, found by integrating 
the gamma-ray cross-section 0.06 cm$^2$ g$^{-1}$ and the net electron mole fraction Y$_e$.\\
The gamma rays are produced from the decay of $^{56}$Ni to $^{56}$Co, releasing photons with an average energy of 1.75 MeV, and $^{56}$Co to $^{56}$Fe, 
releasing photons with an average energy of 3.61 MeV, and positrons with an average kinetic energy of 0.12 MeV.\\
The optical photosphere is defined to be where $\tau$=2/3, integrating inward from the surface. The optical 
opacities are calculated and tabulated using bound-bound, free-free and bound-free transitions
for multi-level atoms of H, He, C, O, Si, Fe. These tables were made for calculations where the H envelope 
was metal deficient, such as SN 1987A. This is not a problem in LC calculations because in the region
where the optical depth is high most of the opacity comes from electron scattering and the metal lines are an insignificant
contribution. In the optically thin regions the opacity is so low that the contribution of metal lines
to the opacity is small. Plus the metalicity of progenitors is unclear, SNe found far from the nucleus in large galaxies 
with steep metalicity gradients most likely will have low metalicity envelopes.\\
The total luminosity (L) at any time is defined as the luminosity 
at the photosphere ($L_{PHOTO}$) plus the energy produced by the deposition of gamma rays above the photosphere 
which is assumed to add to the bolometric luminosity.
\begin{equation}
L = L_{PHOTO}+\int^R_{r_{photo}}m(r)S_{dep}(r)dr.
\end{equation}
Where $r_{photo}$ is the position of the photosphere, $R$ is the radius of the ejecta, $m(r)$ is the mass at radius $r$, 
and $S_{dep}$ is energy per gram per second deposited by gamma rays at radius $r$.
At late times when the gammasphere has receded through the ejecta the energy source of the LC is just the 
spontaneous release of energy deposited by the decay of Ni-Co-Fe.
The absolute magnitude light curves are calculated following \nocite{swh91} Swartz et al. (1991),
\begin{equation}
M_{bol}=-18.793-2.5log_{10}L_{43}
\end{equation}
Where $M_{bol}$ is the absolute bolometric magnitude and $L_{43}$ is the total luminosity in units of $1\times10^{43}$ ergs.
Visual and blue light curves are calculated by assuming a blackbody at the effective temperature integrated
over the V and B response curve given in Azusienis and Straizy (1969) \nocite{as69}. The photospheric temperature, T$_{photo}$, is defined
by 
\begin{equation}
T_{photo}=max\left\{{\left(\frac{L}{4{\pi}R^{2}_{photo}\sigma}\right)}^{0.25},4500\right\}
\end{equation}
following Swartz et al. (1991). Where $L$ is the total luminosity and $R_{photo}$ is the photospheric radius. Finally the velocity at the photosphere, V$_{photo}$ is the velocity of the material at the the photosphere.

\section{Results}
The models A-H (Table I) were constructed in order to easily compare the behavior of one parameter
while holding the other parameters constant. The standard model is model B (M = 16 M$_\odot$, R = 430 R$_\odot$)
with E = $1\times10^{51}$ ergs,
M$_{Ni}$ = 0.07 M$_\odot$, and Ni mixing throughout the 6 M$_\odot$ core. These are similar to the values of SN 1987A 
except with a larger radius, taking into consideration that SN 1987A might have had a nonstandard radius. Thus I take the standard model as  having the most average values around which
the parameters are varied. The 5 parameters are the progenitor 
radius, envelope mass, explosion energy, mass of $^{56}$Ni, and $^{56}$Ni mixing.
All LC graphs are plots of absolute magnitude versus time in days and all calculations 
proceed to 400 days and include the observed LC of SN 1987A for comparison 
\citep{c87}. Figs. 1-3 show the affects of progenitor radius, Figs. 4-6 show the 
affects of ejected mass, Figs. 7-9 show the affects of explosion energy, Figs. 10-12 show
the affects of Ni mass, Figs. 13-15 show the affects of Ni mixing.
Figs. 1, 4, 7, 10, and 13 show bolometric and visual light curves, and photospheric temperature and velocity for 
each of the ``affects of'' series. Figs. 2, 5, 8, 11, and 14, show the density and temperature
profiles for each of the ``affects of'' series at times 0, 36 hours, and 47 days. Also shown is the 
final velocity profile and the luminosity versus mass  at times 94, 189, and 379 days. Figs. 3, 6, 9, 12, and 15, show 
the affects on the LC at different values 
of the parameters that are held constant for each of the ``affects of'' series and for reference compare them to the
instantaneous energy released from the Ni-Co decay. In this way I am 
exploring the parameter space in the most systematic way possible.\\
For example, the affects of radius graphs show models with the
radius ranging from 43 R$_\odot$ to 4300 R$_\odot$ while holding the envelope mass at 16 $M_\odot$, 
explosion energy at $1\times10^{51}$ ergs, mass of $^{56}$Ni at 0.07 $M{_\odot}$ 
and the $^{56}$Ni mixing to 6 $M{_\odot}$. Figs. 1a, 1b, 1c, and 1d show the affects of radius on the bolometric LC, visual LC, T$_{photo}$, and photospheric velocity
 respectively. Figs. 2a, 2c show the affects of radius on the density, temperature
 versus mass at t = 0, 36 hours, and 47 days after explosion. Figure 2b shows the final velocity versus mass and Figure 2d shows the 
luminosity versus mass at t = 94, 189, 379 days after explosion.
Figs. 3a, 3b, 3c, 3d each show the affects of varying the progenitor radius when changing the value of one 
parameter; ejected mass to 8 $M_{\odot}$, 
the explosion energy to $2\times10^{51}$ ergs, the $^{56}$Ni mass to 0.035 $M_{\odot}$, and the $^{56}$Ni mixing $<$
0.3 $M_\odot$, respectively.\\
\section{Discussion}
A general overview of the LC can be explained as follows. 
The LC gets its energy from two sources, deposited shock energy and deposited
gamma rays. Almost always the early light curve ($<$ 50 days) is powered by the diffusive release of 
internal energy deposited by the shock wave as it propagates through the envelope. 
The middle LC (50-120 days), the plateau and secondary peak, is one or a combination of 
both energy sources. This region makes a transition from being powered by deposited shock energy to deposited gamma-ray energy. 
And the late time light curve ($>$ 120 days), the tail, is powered by the instantaneous energy of deposited gamma rays. Throughout the simulation the luminosity is determined by the position of the photosphere calculated by integrating the opacity times the density from the surface inward. As the model expands the opacity and density fall and the photosphere moves inward. This movement of the photosphere toward the center of the model is called the recombination wave (RW).  \\

\subsection{Affects of Progenitor Radius (figs. 1-3)}
In Brief:

\begin{itemize}
\item Larger progenitor radii produce a brighter early LC, and a longer and brighter plateau/secondary peak
\item The progenitor radius has no affect on the late-time light curve
\item The SN light curve of the smaller progenitor radii are dramatically influenced by the affects of Ni
\end{itemize}

The progenitor radius has a significant affect on the early LC and little or 
no influence at all on the late LC. The light curve begins when the shock wave hits the surface. Once the shock breaks out the progenitor radius is the primary variable in determining the luminosity. Larger radii progenitors are already pre-expanded, have a large surface area, and produce a very bright peak (fig. 1a,b). At late times the initial radius has little affect on the gamma-ray deposition.\\
The initial delay in the light curve larger radii progenitors is due to the shock taking longer to arrive at the surface. The shock wave can be identified in figure 2 by the deviation of the long dashed lines and the solid lines for each of the three models.
The time the shock wave spends inside the largest radius progenitor is 8 days compared to 2 hours for the smallest radius.
It can be seen in Figs. 2a, and 2c that the shock in model A 
has already reached the surface by t = 4.2 hours, while both models B and C have shocks moving down the density gradient near 
10 M$_\odot$ and 7.5 M$_\odot$, respectively. In Figure 2c the temperature profiles clearly show a forward shock and a reverse shock forming an mesa in the middle. 
In the largest radii model the deposited shock energy will not do as much PdV work, since it is 
already in an expanded state when the shock arrives, and the material will stay hotter longer.
Once the shock wave emerges for models with increasingly larger radii the radiation has a 
smaller diffusion time and the photon flux is higher accounting for the progressively broader 
initial peak and plateau. A comparison of the diffusion times between the three models clarifies this; 
The ratio of model A to B is 10, model A to C is 100 using diffusion time scale at constant opacity $\sim$ 
3$\kappa$M/(4$\pi$Rc).\\ 
The initial peak of model C has a 4 day rise time compared to the almost instantaneous rise time 
of models A and B (see Fig. 1a, 1b, 1c, 2d). This can be explained by the behavior of the shock
wave in low, constant density material. As the shock nears the surface of model C it becomes smoothed out
and looses it's ``shock'' definition and more mass has a higher temperature, giving a 
broad peak to the photosperic temperature evolution. This is unlike the smaller radii models which have a sharper, 
more defined shock and thus achieves a higher temperature, but then expands and cools faster.\\
The difference between the radii can explain the maximum LC luminosity even though the photospheric temperature in
the largest radii model is lower. The ratio of the largest to smallest radius model is 100, 
while the ratio of the breakout temperature is only about 2.5 (Fig. 1b). This leads to a factor of about 256 brighter in 
luminosity (6 in magnitude, Figs. 1a, 1c, 1d) for the larger radius progenitor.
For smaller stars much more of the internal energy deposited by the shock is 
transformed into PdV work in order to unbind the star. The initial burst is much brighter, seen as a spike, but falls 
rapidly with the expansion. \\
At late times the luminosity profiles are a useful diagnostic for the gamma-ray deposition. All three 
models are shown at t = 94 days (Model A long dash, B dotted, C small dash), and have indistinguishable luminosity profiles
 at t = 189 (solid lines), and 379 (dark solid lines).
The curves at t = 189 and 379 days are understandable since once the ejecta is in homologous expansion the velocity and density
(Figs. 2a, 2b) profiles have similar shapes and thus give similar gamma-ray depositions. The absolute 
magnitude and slope of the LC tails (figs. 1 and 3)
are similar since it is the mass of Ni and the expansion velocities which are important in the deposition of gamma rays.\\
The visual light curve (fig. 1c) is shown for comparison to observations in the V band. The radioactive tails all have a similar
shape to figure 1a but reduced in brightness. This shift is just due to the temperature
being constrained, a temperature floor, when convolving with the V band filters. The temperature floor is set at 4500 K. 
The photospheric velocity (fig. 1d) for the different models show the largest differences quite early and 
converge at about 100 days. After day 35 the V$_{photo}$ is systematically higher for the larger radius 
model because the ejecta is evolving slower and thus the photosphere recedes inward at a slower rate.
In the smallest radius model it is interesting that the increase in the photospheric temperature during the secondary peak does not influence the photospheric velocity. This is due to the photosphere residing in the He core and at temperatures of about 6000 K cannot ionize the He. \\
The recombination wave is an indictor of the dynamics of the temperature and density of the different models. To gauge the dynamics it is instructive to list the times of when each model is completely through the H envelope; smallest to largest progenitor radius, 42, 70, and 104 days (vertical arrows in figure 1b). These models reach the center of the model, completely through the He core, at 60, 103, and 113 days, respectively (vertical lines in figure 1b). \

Figure 3a shows that increasing the explosion energy has an increasing affect on larger radii 
progenitors. The early LC gets progressively brighter as the progenitor radius
is increased because the RW moves inward more rapidly and more of the shock energy gets radiated faster. Thus the characteristic
features of the LC, the peak, plateau, and onset of the radioactive tail, are seen at earlier times. After 200 days the material has expanded enough that gamma rays
have started to escape and the LC begins to show a steeper slope due to the increased velocity
profile of each model being higher than for models presented in figure 1a (see affects of energy section).\\
Figure 3b shows that placing all the Ni at the center delays the affects of the deposited gamma rays and
effectively traps more gamma rays at later times. This has a pronounced affect on the smaller radius 
progenitors since most of the plateau and secondary peak is powered by $^{56}$Ni. Thus the plateau dips to even lower luminosities 
to indicate even more dramatically where the change from shock energy to deposited gamma rays occurs. The secondary peak 
then appears later, flatter, and fainter.\\
Figure 3c shows that reducing the mass of the H envelope produces a light curve that shows features like the onset and duration of the plateau, at earlier times. 
Because of a lower H mass the photosphere can move more 
quickly through the envelope, producing a brighter and faster light curve. The velocity, E/M, is faster than in figure 1a and thus the slope of the tail is steeper. The discrepancy 
of the tails at late times is due to the velocity profile of the He core (fig. 2b). The 8 
M$_\odot$ models with only 2 M$_\odot$ of H envelope have a larger percent of the mass with different velocity profiles
than the 16 M$_\odot$ models. This is the reason why the spreading of
the tails was not seen in the case where the explosion energy was increased to $2\times10^{51}$ ergs (fig. 3a). In that case the velocity profiles were just scaled to higher velocities (also see affects of mass section).\\
Figure 3d shows that the Ni mass affects the smaller radius progenitors more, reducing the plateau dip slightly and 
scaling down the entire secondary peak. The largest radius model is unaffected by the change during the initial peak, plateau, 
and secondary peak. However the luminosity drops further after the secondary peak to meet the tails of the other models. 
The tails also have the same slope as figure 1a only scaled down in absolute magnitude by about 1 magnitude.

\subsection{Affects of Ejected Mass (figs. 4-6)}
In Brief:
\begin{itemize}
\item Increasing the mass of the H envelope causes all the LC features (plateau, secondary peak, and tail) to appear at later times
\item Increasing the H envelope mass produces a fainter early LC and a slowly declining radioactive tail
\item Reducing the H envelope mass changes the final density profiles causing the LC tail to be very steep
\end{itemize}

The general trends seen in Figures 4-6 show that a more massive envelope results in a fainter early LC, a more 
pronounced plateau and secondary peak, and a slower decline of the tail. The affect is due to the 
significant difference in the velocity profiles (fig. 5c). Increasing the ejected mass and keeping the explosion energies the same lowers the 
energy per gram, thus the velocity, and the characteristic features of the LC are seen at a later 
time, a slower LC is produced.\\
The early LC is dependent on the mass because a larger mass envelope will have a longer 
diffusion time and thus trap the radiation longer and results in a fainter peak.
During the plateau region a higher mass H envelope will take longer to expand and cool as seen by the different positions of the RW between the models at day 47 (figs. 5a and 5c). The position of the photosphere at 47 days
(figure 5c) is at the top of RW, located at 12 M$_\odot$ for model B, 8 M$_\odot$ for model E, and 6 M$_\odot$ for model G. 
In the largest mass model the RW has more mass to move through so the photosphere will take 
longer to arrive at the He core, accounting for the longer plateau duration. 
Figure 4b supports this showing that the photospheric temperature for larger ejected mass stays hotter for a longer time.
The smaller the mass of the envelope the higher the velocity profile (fig. 5b), the lower the
density profile (fig. 5a), the lower the temperature profile (fig. 5c), the faster the RW and thus 
little time for the plateau to form. 
Since the late time LC is dependent on the trapping of gamma rays the smaller mass 
explosions will start losing gamma rays sooner for two reasons. First the number of absorbers is reduced
and second the density profile is lower at any given time. Thus the LC tail will have an 
increasingly steeper slope as seen in figs. 4a, 4c, and 4d. Figure 4a shows the affect of an extreme case of having no hydrogen envelope. This model has only the 6 M$_\odot$ He core, with its original unmodified radius. The Ni mass and Ni mixing are consistent with the other three models. It can be seen that that the initial rise in brightness is reduced since the H envelope mass is absent and the progenitor radius is smaller. The affect of removing the H envelope causes the LC to have an earlier secondary peak similar to that seen from explosions of Wolf-Rayet stars \citep{ew91}. The secondary peak and the LC tail are not as bright compared to the other models due to the photophere receding faster into the center of the model and to the fewer absorbed gamma rays heating the material and adding to the luminosity.
The visual LC's (fig. 4c) show that for a smaller ejected mass the initial peak becomes more defined,
whereas the higher mass models show a more extended plateau.
The photospheric velocity (fig. 4d) for all three models have identical early fall times, 
showing that the expansion in the outer regions is very similar. Then after about 25 days they 
start to separate showing that as the photosphere reaches the He core and the photosphere recedes 
to much lower velocities faster. For the 8 M$_\odot$ model this occurs at 40 days, 12 M$_\odot$ model at 65  days,and  16
M$_\odot$ model it is 90 days.
The influence of the heating due to gamma rays can be seen in the luminosity profiles in figure 5d. 
At t = 94 days (solid lines) the 8 and 12 M$_\odot$ models have
the same luminosity profiles indicating that the photosphere has receded into 
the most inner most material and the total luminosity is given by the 
trapping of gamma rays. Due to its large mass the 16 M$_\odot$ model can  still 
release energy stored from the trapped gamma rays out past day 94. At t = 187 days (dashed 
lines) the photosphere in all models has receded to the inner most mass zones. The 
8 M$_\odot$ model has had a movement of the gammasphere inward in mass 
indicated by the the lower luminosity profile. This is evidence that the 
material has become thin to gamma rays and some are escaping. By t = 379 days
(solid dark lines) the luminosity profiles are diverging, showing that the 
smaller mass models have increasingly more gamma rays escaping. \\
Figure 6a shows that for a higher energy the early LC is just scaled brighter due to the increased shock 
energy being released. The plateau is shorter since the material is expanding faster and thus
cooling faster and the recombination wave moves inward in mass faster. The most pronounced affect is 
the fanning of the tails due to the increased expansion velocity and thus the escape of gamma rays.
This shows that increasing the explosion energy has an increasing affect on the escape of gamma rays 
for smaller masses.\\
Figure 6b shows that with smaller amounts of Ni the early LC is unchanged, but the luminosity drops near the end of the plateau, 
affecting both the secondary peak and tail. The diffusive release of shock energy was not influenced by the 
reduction of Ni mass so no change in the early LC or most of the plateau is seen. The secondary peak is affected
since it is partially powered by Ni and figure 6c shows a reduction in its brightness and width.
The most significant affect is the tails which show the same slopes as in fig 3a but the absolute 
magnitude is greatly reduced.\\
Figure 6c shows that for a larger radius the early LC is much brighter due to a larger radiating surface.
The duration of the plateau is increasingly longer for larger ejected masses. However the width
of the secondary peak doesn't seem to be influenced at all. This is due to the recombination
wave moving through the He core which doesn't have the required internal energy to sustain the higher luminosity. 
Thus it recedes fast releasing stored shock energy and deposited 
gamma rays. The tail slopes and absolute magnitudes are exactly the same as for figure 3a (see Affects of Radius).\\
Figure 6d shows that confining the Ni to the innermost mass layers ($<$ .3 M$_\odot$) has an increasingly
larger influence on the smaller ejected masses. For smaller masses the secondary peak is much
longer in duration due to the appearance of the large dip at 25 days. 
The dip is also found for larger masses but the affect is reduced.  
The reason for the large dip is due to the change in the LC being powered by deposited shock energy to deposited 
gamma rays. This transition between energy sources is a smooth transition, possibly due to the treatment of gamma-ray deposition. Other light curve studies have used the mixing of H and He to reproduce the same transition for SN 1987A (Woosley 1988, Shigeyama 1988, Utrobin 1991). The purpose of H/He mixing was to reduce the number of free electrons and thus expedite the recombination wave inward. The photosphere then reaches the Ni bubble faster and the affect on the light curve is similar to mixing the Ni outward, showing a dip and a well defined secondary peak. 
In both scenarios the photosphere rapidly reaches the Ni bubble where the material is hot enough to cause an increase in the luminosity.
At late times the tails all show total trapping of the gamma rays.
The affects of mass shown here are consistent with Woosley (1988), Shigeyama and Nomoto (1990), 
Arnett (1989), and Hsu et al. (1992) when comparing both the luminosity differences in the secondary peak
 and time of the secondary peak maximum.

\subsection{Affects of Explosion Energy (figs. 7-9)}
In Brief:
\begin{itemize}
\item Higher explosion energies produce a LC with features (onset and duration of plateau, secondary peak) that occur at earlier times
\item Higher explosion energies produce a bright early LC and a steeper slope of the tail
\item Increasing the energy does not change the density profiles as significantly as changing the envelope mass (section 5.2), limiting the spread in the radioactive tails
\end{itemize}

The general trends found when the explosion energy is varied can be seen in Fig 7a. As the energy is increased the early LC 
and the plateau are scaled brighter and the tail has a steeper slope.
Increasing the explosion energy produces a larger E/M, energy per gram, and thus a higher overall
velocity profile. The higher the explosion energy the more energy deposited, the higher the temperature 
and the brighter the early LC and plateau in most cases because they are powered by the release of 
deposited shock energy. The larger velocities expand and cool the material giving 
a faster RW and and a faster overall LC. At later times the LC tails
for a higher explosion energy have steeper slopes, but not as much of an affect as when varying the mass  
(see Affects of Mass). This can be understood from a comparison of velocity profiles. The velocity profile (fig. 8b) 
shows that the final velocities do not differ by nearly as much as compared with figure 5b (varying the
M$_{env}$). Furthermore the density profiles figure 8a and temperature profiles figure 8c show more
similar profiles for different explosion energies than figures 5a and 5c. 
At late times the ``affects of energy'' series of models have very similar density and velocity profiles in comparison to the ``affects of mass''series of models. The similarity in velocity and density profiles accounts for the similarity in the LC tails in figure 7a. 
The important result here is that given a certain E/M it does not necessarily mean varying E or M$_{env}$ will result in the same LC 
tail.\\ A comparison between changes in E or M$_{env}$ also affects the photospheric temperature of the light curve.
The photospheric temperature (fig. 7b) shows a much slower response to a variation in energy than the photospheric temperature 
when varying the mass (fig. 4b). A comparison in the time difference 
of when the photospheric temperature drops to 4500 K between varying the energy and varying 
the mass shows this. When the energy is tripled the time difference is only 27 days as opposed to when the mass is 
just halved the difference is 57 days. Thus the temperature of the photosphere is much more 
sensitive to a change in mass rather then a change in energy.\\
The visual LC (fig. 7c) shows very little or no change in the LC shape between differences in models when compared to figure 1a, except for the initial peak becoming part of the plateau.
The photospheric velocity for all models is relatively the same until the photosphere enters the 
He core region where the velocities separate. The models with the highest energy show that they reach 
the He core the fastest and thus start to move into the slower moving material earlier.
The luminosity profiles figure 8d show that for the least energetic model (long
dashed lines) the affect of the gamma-ray heating wave is pushed to later 
times. The more energetic models have already passed through that stage by the same time period. By t = 190 days the profiles are exactly
the same (solid lines) indicating that the difference in energies hasn't 
changed the deposition of gamma rays. At t = 379 days the differences in the 
deposition can start to be seen as the luminosity profiles start to separate slightly for the different models.\\
Figure 9a shows that a smaller the ejected mass results is a smaller difference in the early LC and a steeper fall of the tail.
This is understandable since the smaller the H envelope the faster the recombination wave and  
the faster the release of internal energy deposited by the shock. At late times the steeper tail can be explained by 
a higher overall velocity profile than figure 6b. This is due to the lowered H mass, and progressively more gamma rays 
escape with increasing explosion energy.\\
Figure 9b shows that when Ni mixing is changed to $<$ 0.3 M$_\odot$ the early LC is exactly the same as in fig 1a. But the plateau and secondary peak are systematically affected more. There is a clear separation between the thermal and Ni energy. This is seen as a more defined plateau/secondary peak. The tails are nearly the same since most of the gamma rays are absorbed.

Figure 9c shows that changing the radius to 4300 R$_\odot$ distinguishes between the LC peak times which are about 1.4 and 1.7 times earlier for the $2\times10^{51}$ and $3\times10^{51}$ 
erg models respectively. The affects of a larger radius show a narrower range in the 
duration, slope, and absolute scale of the plateau as 
compared to the same graph comparing ejecta masses (fig. 6c). The secondary peak is smaller in width and starts to blend 
with the plateau. This is due to more shock energy deposited into the H envelope and He core.
The tail is almost identical to figure 7a but with a greater change in slope with increasing energy. Again the influence is much smaller than with a variation in mass (fig. 6c).\\
Figure 9d  shows that when the Ni mass is decreased the secondary peak gets increasingly fainter but more importantly the 
the width of the peak gets smaller with increasing energy. This means that as the explosion energy gets higher, 
the secondary peak is more powered by gamma-ray energy deposition, while the lower energy explosions have a longer diffusion time and the 
shock energy can still contribute to the secondary peak. At later times the slope of the tails are similar to
figure 7a but scaled to lower absolute magnitudes.\\

\subsection{Affects of $^{56}$Ni Mass (figs. 10-12)}
In Brief:
\begin{itemize}
\item Larger $^{56}$Ni mass produces brighter radioactive tails and plateaus/secondary peaks
\item Models that have smaller radii are influenced more dramatically by the $^{56}$Ni mass
\item Larger E/M and extensive $^{56}$Ni mixing reveals the $^{56}$Ni affects earlier and steeper slope of the tail
\end{itemize}

The general trend found when varying the $^{56}$Ni mass shows that as the $^{56}$Ni mass is increased the plateau, secondary 
peak and tail of the LC get brighter.
 The early LC is powered by deposited shock energy
so changing the Ni mass should not affect this part of the light curve. By about 20 to
50 days the photosphere moves into the region where gamma-ray deposition is 
important. The affects of Ni mass are apparent by the difference in absolute magnitude and shape of the LCs. 
In figure 10a the affects of an increased gamma-ray deposition can be seen by all three LC's
diverging at $\sim$ 40 days. As the Ni mass is
increased the secondary peak increases in both absolute magnitude and width. In fact when the Ni mass is 0.14 M$_\odot$
the secondary peak is brighter than the plateau, and in the V light curve 
the secondary peak is even brighter than the initial peak (fig. 10c). At late times the V light 
curve have similar tail slopes but scaled to lower luminosity (equation 3). For comparison a model is shown that contains no Ni mass and abrutly falls after the plateau. Since having no Ni mass eliminates the secondary peak it is reasonable to assume that Ni heating plays a role in producing the secondary peak (fig 10a).\\ 
The plateau is actually lengthened by increased amount 
of Ni as seen in all graphs to the point where it is almost doubled for 0.14 M$_\odot$ Ni. The time of the secondary peak maximum
increases with increasing Ni mass. This can be explained by the Ni-Co decay keeping the material 
hotter for a longer time, thus slowing the RW. Figure 10b shows the photospheric temperature for the 0.14 M$_\odot$ Ni model stays hot for 
greater than 115 days, indicating that the gamma rays participate in the heating of the material below
the photosphere. There is little change in the photopspheric velocity (fig. 10d) which shows no difference until about 55 days. The slight difference in velocity after 55 days is due to an increase in the opacity, causing the photosphere to move into faster moving material. The slopes of all the tails (figs. 10a, 10c) are similar since there is
no variation in mass, energy, or Ni mixing. This can be easily explained since all models have almost identical density,
velocity, and temperature profiles figs. 11a, 11b, 11c.
In figure 11c the heating due to Ni can be seen to slightly affect the temperature profile at day 47 in the 
region $<$ 6 M$_\odot$.
Figure 11d shows the most predominate affect of Ni mass. The luminosity profile directly reflects the amount of energy supplied by radioactivity, 
similar profiles but different absolute luminosities. In general since the only source of energy at lates times is the decay of Co it is possible to estimate the Ni mass based on the absolute magnitude of the tail. However, a LC tail with a steeper slope than the decay rate of Co indicates that gamma rays are escaping the ejecta and thus an under-estimate of the Ni mass would be obtained.\\

Figure 12a shows that lowering the H envelope mass enables Ni to power the full plateau. The affects of Ni appear
at day 20 due to the recombination wave moving quickly through the low mass H envelope and uncover the 
regions heated by gamma rays faster. As the Ni mass is increased the He core is kept hot
enough to allow the RW to move more slowly. At later times the tails fall faster, when compared to figure 10a, due 
to the faster velocity, low density and thus less trapping of gamma rays.\\
Figure 12b shows that confining the Ni to $<$ 0.3 M$_\odot$ delays the affects of the Ni energy source and consequently the LC continues to fall 15 days longer than the other models in Figure 10a. This is due to the Ni taking longer 
to diffuses through more material. It also produces a 
broader and longer secondary peak which also lengthens the plateau. This is due to the Ni being trapped longer and keeping
the material hotter and thus keeping the RW from moving inward too quickly.\\
Figure 12c shows that increasing the radius lets the deposited shock energy power the LC for a longer time, 
since the affects of the Ni mass aren't seen until 80 days (whereas for fig. 10a it is 40 days). 
Varying the Ni mass 
doesn't have as much of an affect on the plateau or the secondary peak because the RW travels inward 
slower due to the longer diffusion time.
Just befor the secondary peak the photosphere moves
through the He core the LCs start to diverage at about 140 days, showing that the model with the largest Ni mass 
stays brighter for a longer time.\\
Figure 12d shows that increasing the explosion energy, like decreasing the mass, lets the gamma rays power the LC 
earlier and lets the gamma rays escape at later times, so the LC falls below the instantaneous Co slope.
Again the affects of increasing the energy are not as dramatic as lowering the H mass by the same factor.

\subsection{Affect of $^{56}$Ni Mixing (figs. 13-15)}
In Brief:
\begin{itemize}
\item Extensive $^{56}$Ni mixing causes the Ni heating to be seen only slightly earlier and the plateau to be slightly brighter
\item Confining the $^{56}$Ni delays and enhances the heating when the photosphere reaches the  Ni-powered diffusion wave (Ni bubble)
\item Larger $^{56}$Ni mixing causes only a slightly steeper radioactive tail
\end{itemize}

The general trends found when the Ni is mixed further out in the ejecta the earlier the 
affects of the gamma rays are seen and at late times the slope of the tail becomes steeper. However the affects of Ni mixing are not
as dramatic as varying the other parameters. Figure 13a shows that if Ni is mixed further out the 
gamma rays have a longer path length and deposit their energy
at greater distances. So as the photosphere recedes, the more mixed models show the 
influence of the gamma rays earlier, see figs. 13a, 13b, 13c. It is expected that as mixing of Ni is varied
there should be a time difference between models as to when the Ni starts to influence the light 
curve. The model with the greatest Ni mixing (11 M$_\odot$) has a LC that becomes brighter  earlier, by about 20 days, than the other two less mixed models. This is expected since the photosphere will encounter the gamma-ray heated material faster in the 
model with greater Ni mixing.\\
The more mixed models have a brighter and slightly broader plateau due to a combination of two 
things. At earlier times the plateau is brighter because more gamma-ray energy depostions can heat the material nearer to the photosphere adding to the luminosity.  Secondly gamma-ray
deposited energy outside the photosphere adds directly to the luminosity. 
The photospheric temperature of the more mixed models is larger but not enough to account 
for the difference in 
brightness on the plateau, thus the additional luminosity is due to gamma ray deposition outside the photosphere.
The secondary peak becomes broader as the Ni mixing is more confined to the central region. This is 
due to the slow diffusion of material being heated by the gamma rays, the Ni bubble or Ni-powered diffusion wave. 
The photosphere can be seen to increase in temperature as the RW reaches the Ni bubble in figure 13b and is 
shown to be slightly earlier for more mixed models.
For the more mixed model this produces a wider plateau that merges with the secondary peak. For the least mixed model
the plateau ends earlier and produces a dip in the light curve, resulting in a more defined secondary peak (Figs. 13a,c). 
Overall the Ni mixing is not 
a large factor in differentiating models.
In Figure 14d the more mixed models 
have a luminosity profile that increases with increasing mass, showing the location of the Ni distribution. At later times the more mixed models have gamma rays escaping, as seen by
a scaling down of the luminosity profile, indicating that the gammasphere is moving inward faster. 
The density, 
temperature or velocity profiles (figs. 14a, 14c, 14d) show very little difference for all models out to 400 days. \\

Figure 15a shows that for a smaller ejected mass the Ni mixing between the core (dotted line) and the envelope (dashed line) is 
not significant, which is expected since it is only a difference in mixing of 1 M$_\odot$. However 
the least mixed model (solid line) shows a dramatic early drop in luminosity with very little plateau. As the 
photosphere recedes into the He core material that has been heated by the trapped gamma rays it 
produces a secondary peak similar to SN 1987A. In the least mixed model since all
the gamma rays are trapped the radioactive LC tail follows the instantaneous Co decay slope.\\ 
Figure 15b shows that when the amount of Ni is reduced the plateau drops earlier and the light curves 
look more similar to each other. Compared to figure 13a the secondary peak arrives earlier, has a lower absolute magnitude 
and a shorter width. And as expected the difference in the tails is 
similar to figure 13a but scaled fainter. These light curves represent the most severe case of 
no affect on the LC.\\
Figure 15c shows that with an increased progenitor radius the mixing of Ni has very little affect on any part of the
light curve. \\
Figure 15d shows that when the explosion energy is increased the early LC become both more similar and 
brighter, and the tails tend to diverge only slightly more than figure 13a. This indicates that at higher energies Ni mixing is even less important. 
\section{Conclusion}
A numerical parameter study of light curves showing the affects of five parameters representing the progenitor and explosion
is carried out to 400 days. In this parameter study it is found that the early LC is affected by R, M, and E; the plateau 
is affected by R, M, E, M$_{Ni}$,and $^{56}$Ni mixing;
the secondary peak is affected by the R, M, E, M$_{Ni}$, and $^{56}$Ni mixing; and the tail is 
affected by M, E, M$_{Ni}$, and $^{56}$Ni mixing. It was found that the primary parameters that 
influence the overall behavior of the LC are the progenitor radius and the ejected mass. These 
are the two parameters that give the largest changes in the LC with reasonable variations in the 
parameters. The secondary 
parameters are the explosion energy, Ni mass, and Ni mixing. The ejected mass was determined a primary 
parameter since the ejected mass is thought to vary from 4 M$_\odot$
to up 25 M$_\odot$ and affects on the light curve sufficiently account for the observed LC variation. The explosion 
energy is taken as a secondary parameter since the explosion mechanism is not well determined and is thought 
to span a smaller range in parameter space for normal type IIs, from 0.5 to 2$\times10^{51}$ ergs.\\
It is possible that a given LC is not unique since many variables affect the shape and absolute 
magnitude of the light curve. However it is more plausible to say that there is a small range in 
parameter space where the LC is not unique. It is then necessary to look at the entire
LC (out to 400 days or later) in order to determine accurate values of the parameters. It is also beneficial to have a consistent model by fitting the spectrum in conjunction with the light curve, similar to the analysis of SN 1997D by \cite{t97}.\\
The progenitor radius has the largest affect on the early LC but it was also varied by the largest factor, 
100 compared to a factor of 3 for the explosion energy. The radius is the one parameter that can 
reasonably explain the range in peak absolute magnitudes of the light curve. \\
The conspicuous secondary peak found in almost all light curves just after or in place of the plateau is 
due to the recombination wave moving quickly through the compositionally unmixed He core and reaching the area where gamma-ray energy was deposited.\\
The influence of $^{56}$Ni becomes increasingly important for stars with smaller 
radii and smaller ejected mass or larger explosion energy. But $^{56}$Ni has its most significant 
affect on a smaller radius as demonstrated with SN 1987A. Thus as the progenitor radius 
is decreased the plateau and secondary peak become powered by the deposition of gamma rays. The affect of Ni mixing on the light curve can dissapear if the progenitor radius is large or the amount of Ni in the ejecta is very small.\\
The variation of ejected mass does not necessarily show inverse affects when compared to changing the explosion energy. This is 
especially true for the plateaus, secondary peaks, and tails. It is 
shown that changing the H envelope mass causes differences in the density and velocity profiles that change the light curve more dramatically than the explosion energy. Thus it
would be expected that if the He core and the H envelope mass were changed in proportion the inverse 
affects should appear. It is also expected that if polytropes were used then changing the 
envelope mass would give more similar inverse affects.\\

\acknowledgments
I would like to sincerely thank Eddie Baron and David Branch for resources, guidance, and necessary discussions, Robert Harkness for the original hydrocode and opacity tables, and Stan Woosley for the numerically evolved stellar models. I am grateful for the referee whose extensive comments and suggestions greatly improved the paper. This work was supported under NSF grant EPS-0132289 and NASA grant NCC5-399.

\begin{figure}
\plotone{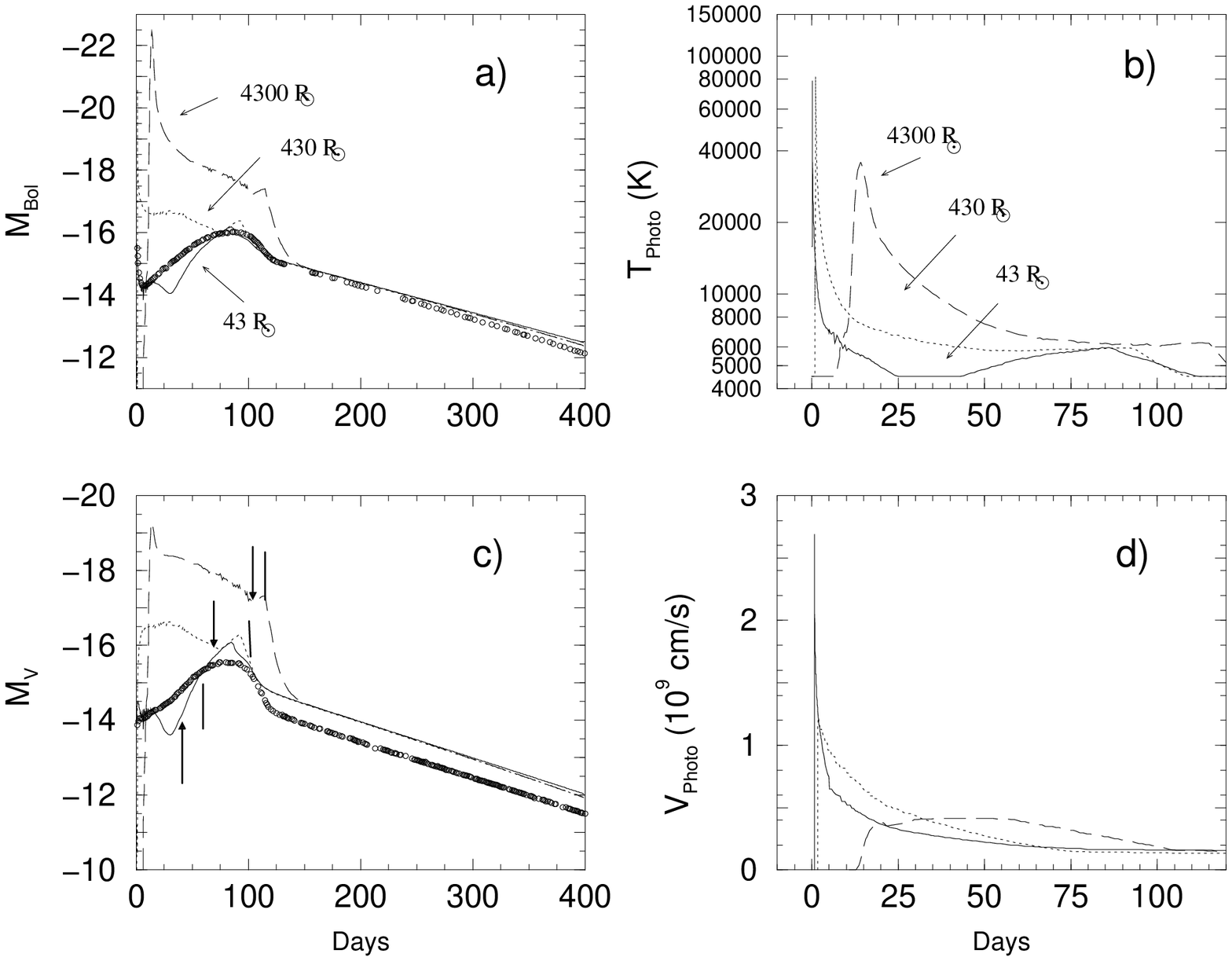}
\caption[Figure 1]{Illustrates the affect of progenitor radius on light curve, photospheric temperature and velocity
versus time for models A (dashed line), B (dotted line), and C (solid line). All models have 
the same total mass, energy, Ni mass and Ni mixing: M = 16 M$_\odot$, E = $1\times10^{51}$
 ergs, M$_{Ni}$ = 0.07 M$_\odot$, Ni mixing throughout the He core. The plots a) and c) show the 
absolute magnitude bolometric and visual light curves of each model compared to SN 1987A, respectively. The markings in plot c) show the position of the recombination wave at the H/He boundary (vertical arrow) and He/C boundary (vertical line). Plots 
b) and d) shows the photospheric temperature and velocity of each model versus time, respectively.}
\end{figure}

\begin{figure}
\plotone{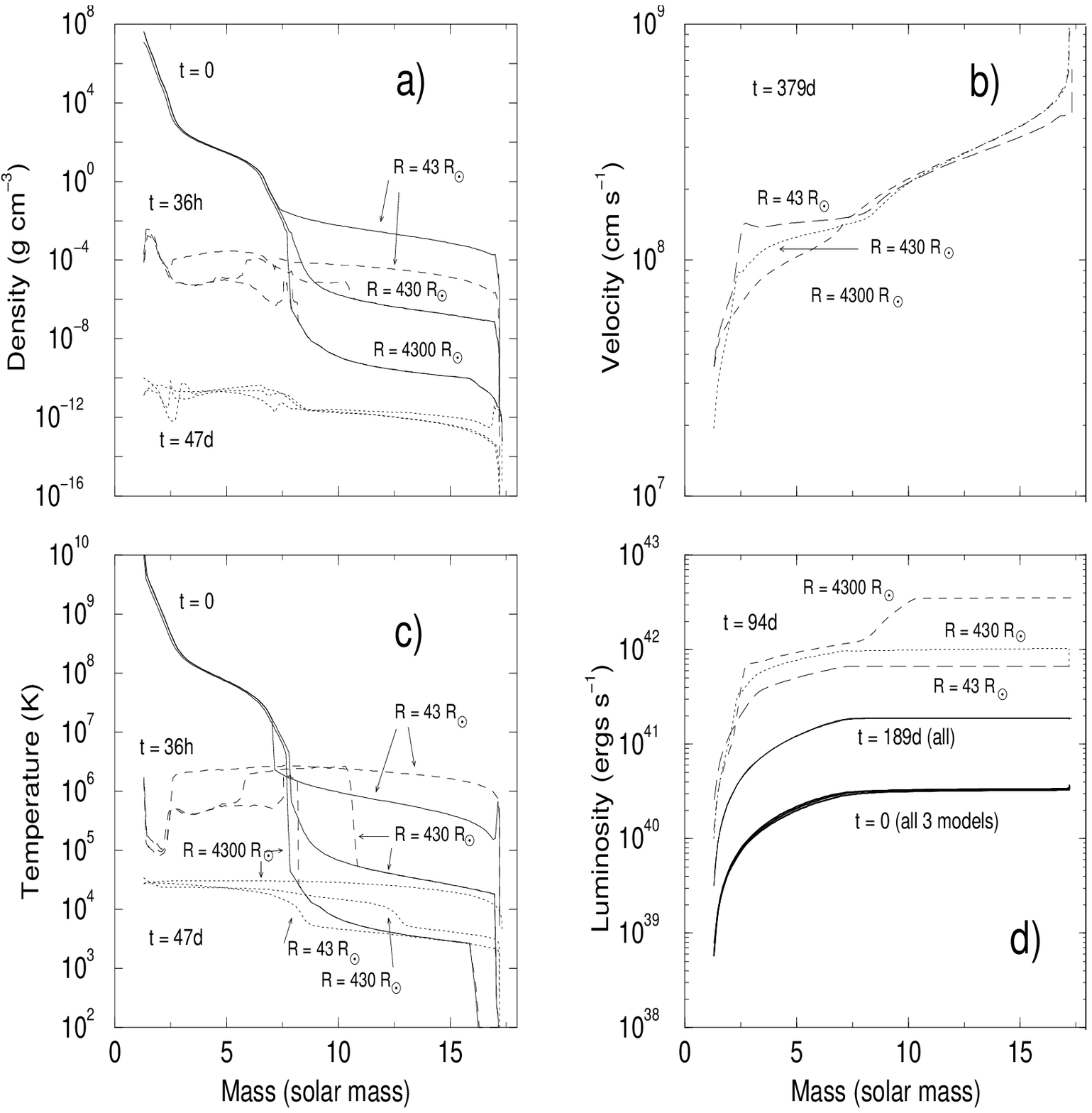}
\caption[Figure 2]{$\rho$, v, T, and L versus mass for the three models in figure 1 are examined at different 
times. Plots a) and c) show the density and temperature versus mass at time = 0 (solid line), 36 hours 
(dashed line), and 47 days (dotted line). The shock can be seen at time = 36 hours in both the 
density and temperature profiles. The recombination wave can be seen at time = 47 days in the 
temperature profile. Plot b) shows the final velocity profile for all three models. Plot d) shows the 
luminosity versus mass at time = 94 days, 189 days, and 379 days. The Ni heating supplies all of 
the energy by day 189 and shows no difference in the luminosity profiles through day 379.}
\end{figure}

\begin{figure}
\plotone{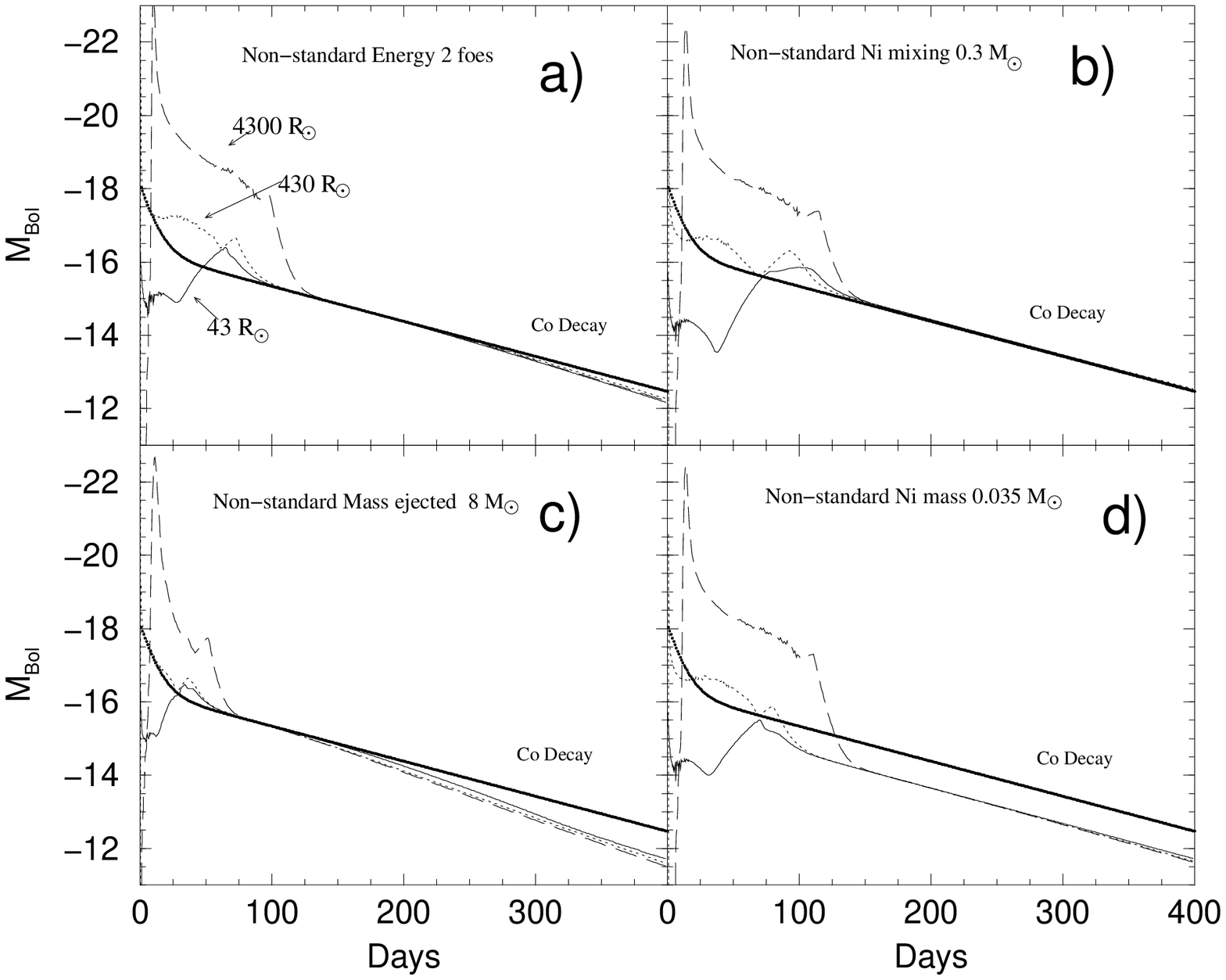}
\caption[Figure 3]{Absolute magnitude bolometric light curves showing the variations of the progenitor radius with different constant values of the parameters. For 
reference the spontaneous luminosity of the Ni-CO-Fe decay for 0.07 M$_\odot$ Ni is shown. Plot a) Shows the light curves of 
the same models as in figure 1 except the explosion energy for all three models is $2\times10^{51}$ 
ergs. Plot b) Shows the light curves of the same models as in figure 1 except the Ni mixing for all 
three models is $<$ .3 M$_\odot$. Plot c) shows the affect of progenitor radius except the total mass 
is reduced to 8 M$_\odot$ (models F, G, H). Plot d) Shows the light curves of the same models as in 
figure 1 except the Ni mass for all three models is reduced to 0.035 M$_\odot$.}
\end{figure}

\begin{figure}
\plotone{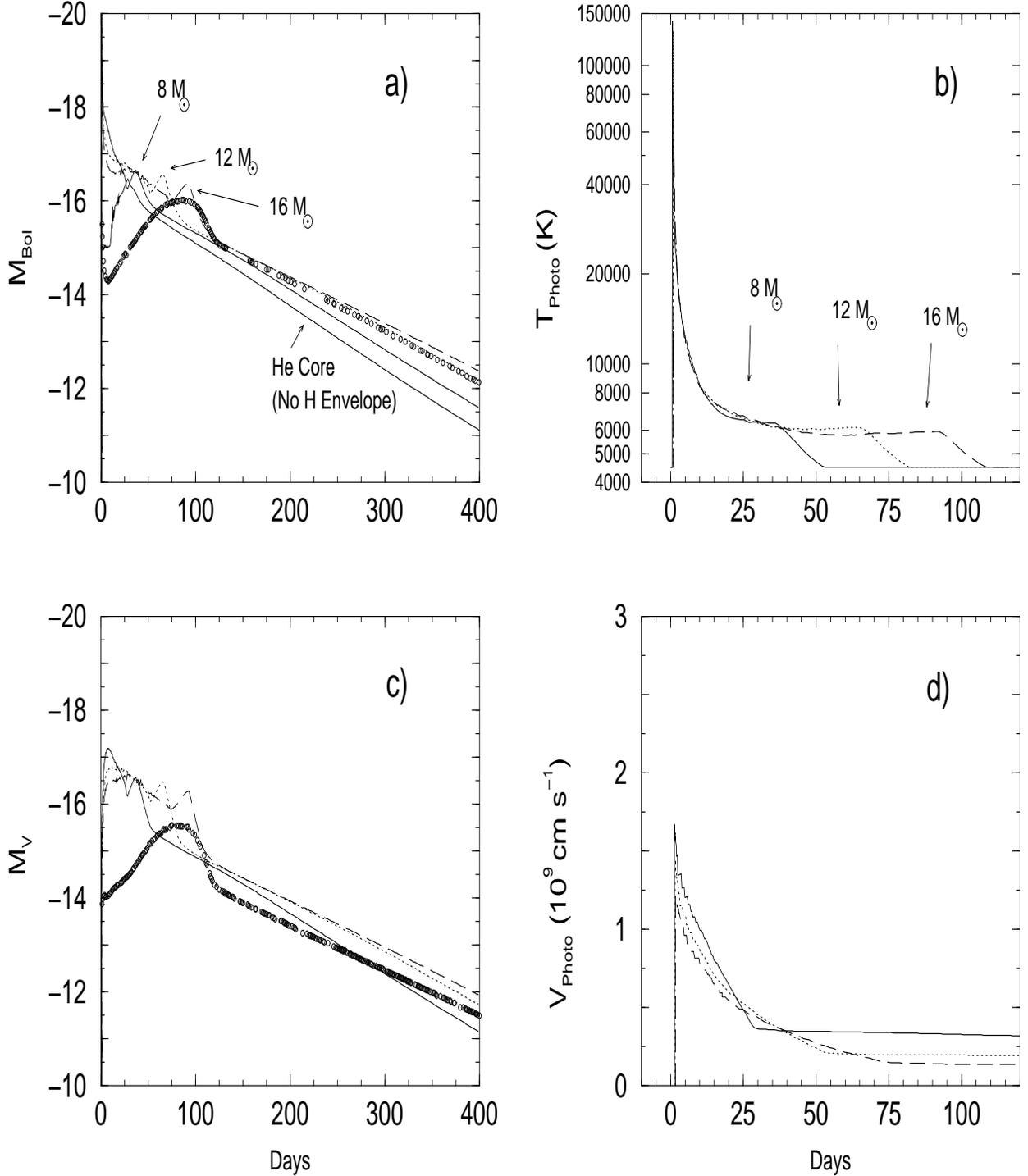}
\caption[Figure 4]{Illustrates the affect of the total ejected mass on light curve, photospheric temperature and velocity
versus time for models B (dashed line), D (dotted line), and G (solid line). All models have 
the same progenitor radius, energy, Ni mass and Ni mixing: R = 430 R$_\odot$, E = $1\times10^{51}$
 ergs, M$_{Ni}$ = 0.07 M$_\odot$, Ni mixing throughout the core. The plots a) and c) show the 
absolute magnitude bolometric and visual light curves of each model compared to SN 1987A, respectively. Plots 
b) and d) shows the photospheric temperature and velocity of each model versus time, respectively.}
\end{figure}

\begin{figure}
\plotone{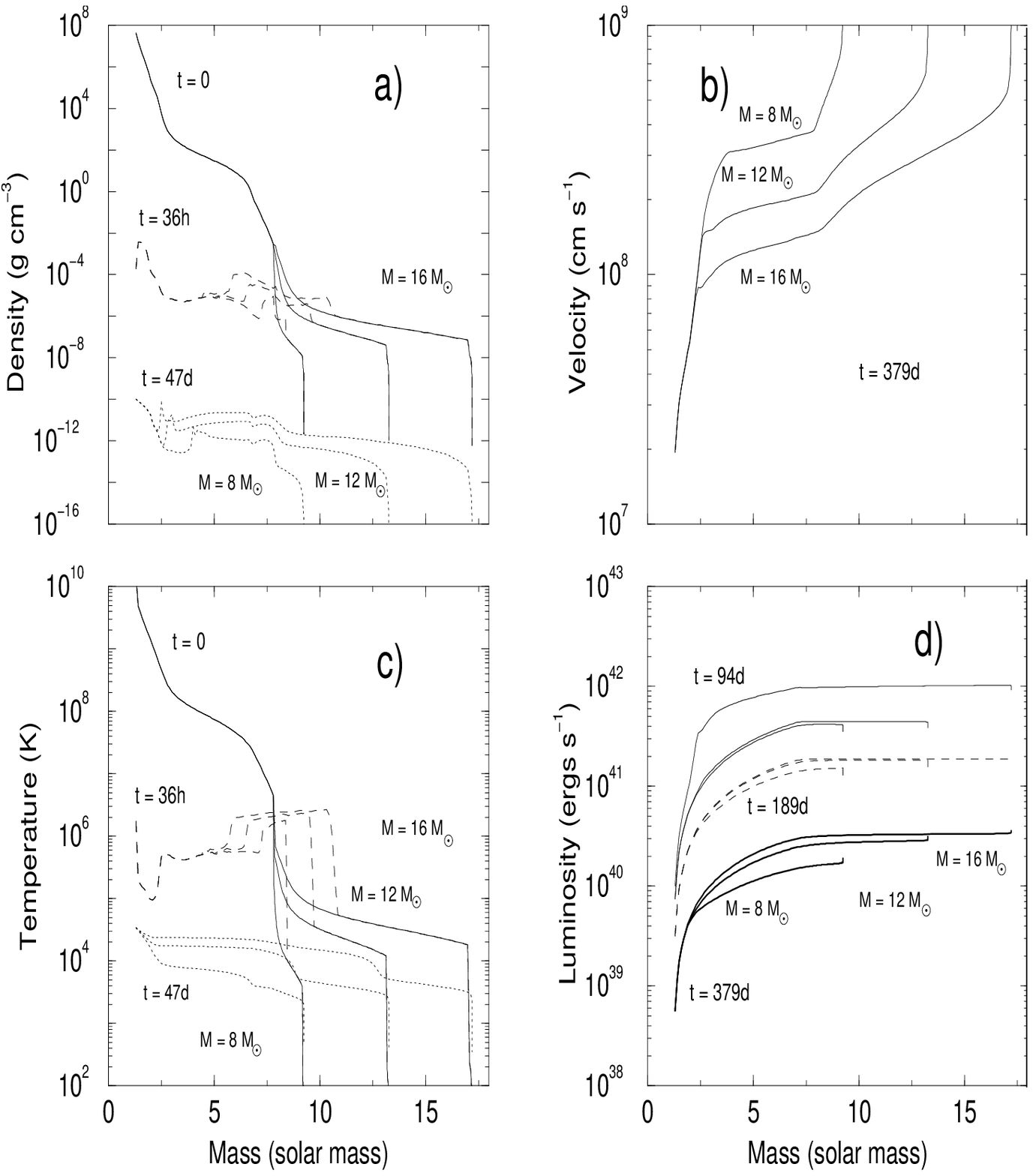}
\caption[Figure 5]{$\rho$, v, T, and L versus mass for the three models in figure 4 are examined at different 
times. Plots a) and c) show the density and temperature versus mass at time = 0 (solid line), 36 hours 
(dashed line), and 47 days (dotted line). The shock can be seen at time = 36 hours in both the 
density and temperature profiles. The recombination wave can be seen at time = 47 days in the 
temperature profile. Plot b) shows the final velocity profile for all three models. Plot d) shows the 
luminosity versus mass at time= 90 days, 189 days, and 379 days. The Ni heating supplies all of 
the energy by day 189 and the luminosity profiles separate by day 379 indicating a loss of gamma rays.}
\end{figure}

\begin{figure}
\plotone{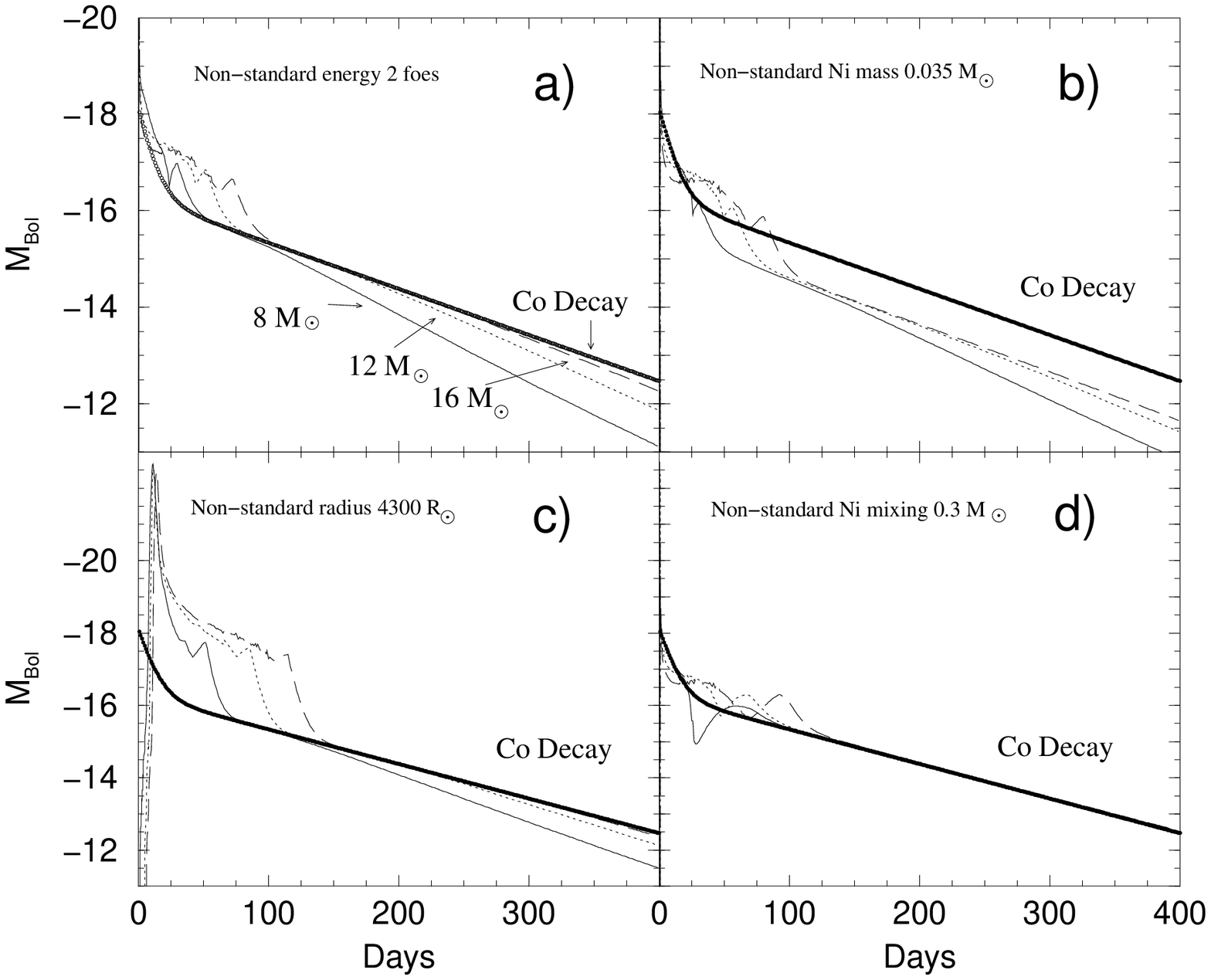}
\caption[Figure 6]{Absolute magnitude bolometric light curves showing the variations of the ejected mass with different constant values of the parameters. For 
reference the spontaneous luminosity of the Ni-CO-Fe decay for 0.07 M$_\odot$ Ni is shown. 
Plot a) Shows the light curves of the same models as in figure 4 except the explosion energy for all three models is $2\times10^{51}$ ergs. 
Plot b) Shows the light curves of the same models as in figure 4 except the Ni mass for all three models is reduced to 0.035 M$_\odot$.
Plot c) shows the affect of ejected mass with the progenitor radius is increased to 4300 R$_\odot$ (models C, E, H). 
Plot d) Shows the light curves of the same models as in figure 4 except the Ni mixing for all three models is $<$ .3 M$_\odot$.}
\end{figure}

\begin{figure}
\plotone{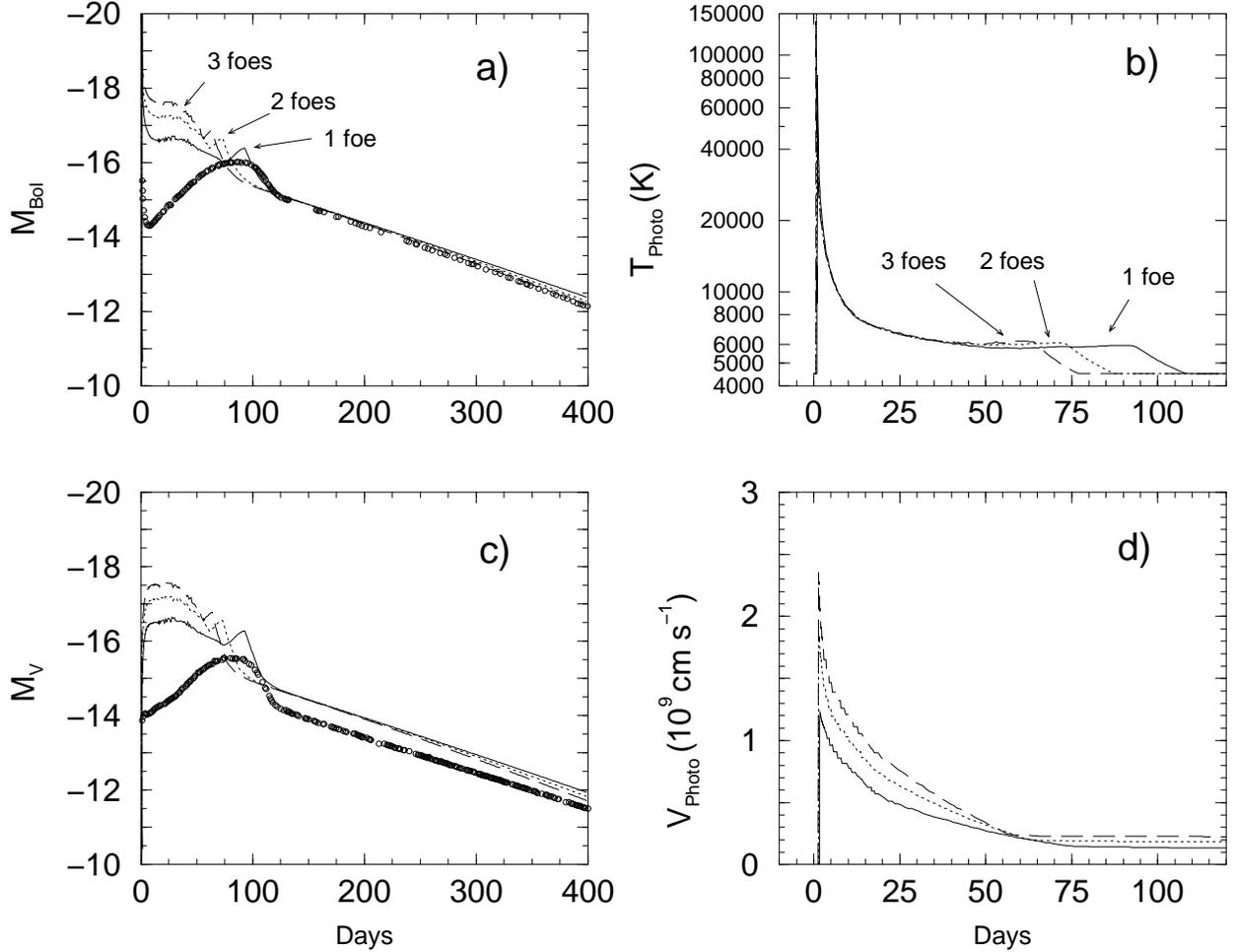}
\caption[Figure 7]{Illustrates the affect of the explosion energy on light curve, photospheric temperature and velocity
versus time for models B with $3\times10^{51}$ ergs (dashed line), B with $2\times10^{51}$ ergs (dotted line), 
and B with $1\times10^{51}$ ergs (solid line). All models have 
the same progenitor radius, ejected mass, Ni mass, and Ni mixing: R = 430 R$_\odot$, M = 16 M$_\odot$, M$_{Ni}$ = 0.07 M$_\odot$, Ni mixing throughout the core. The plots a) and c) show the absolute magnitude bolometric and visual
light curves of each model compared to SN 1987A, respectively. Plots b) and d) show the photospheric temperature 
and velocity of each model versus time, respectively.}
\end{figure}

\begin{figure}
\plotone{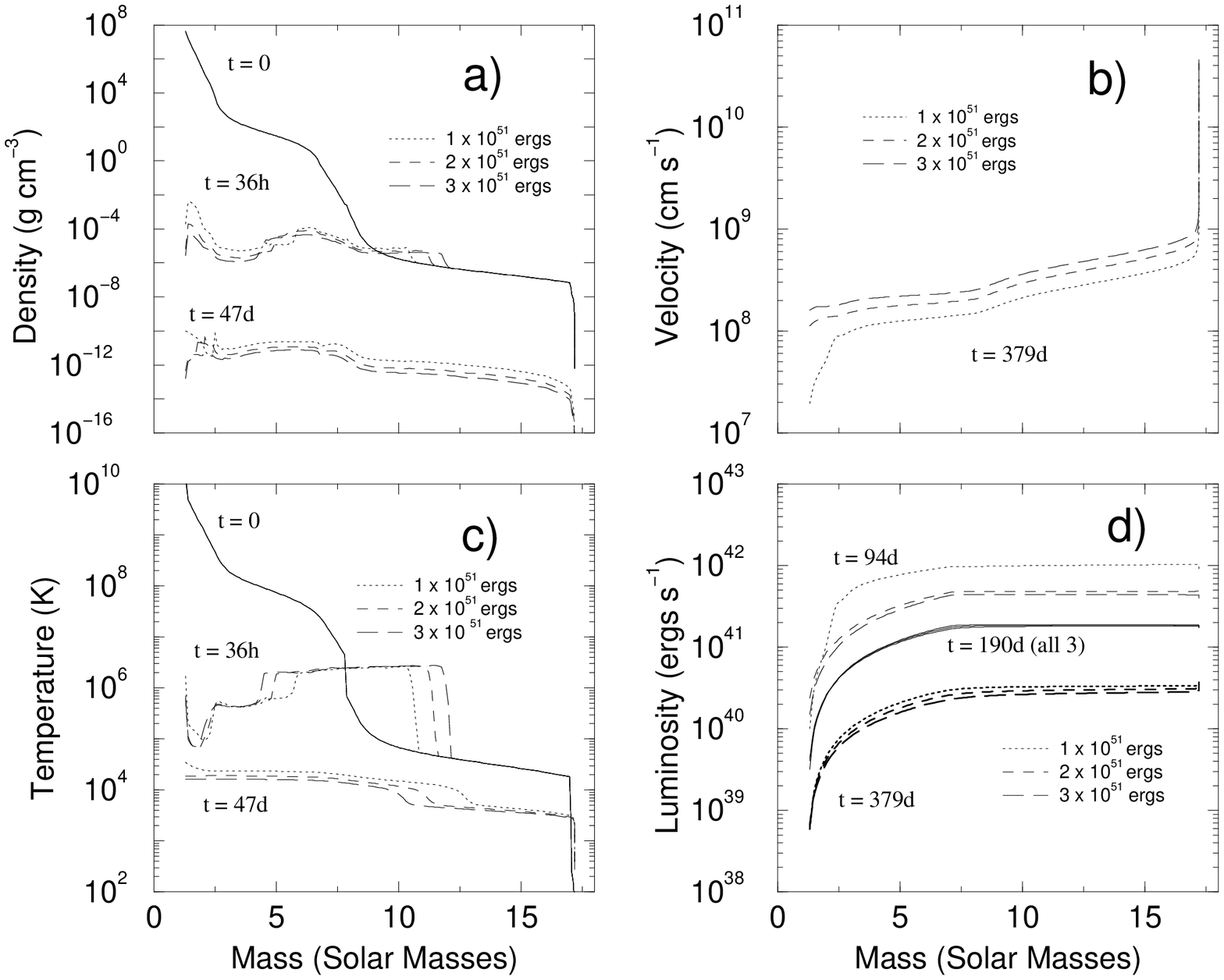}
\caption[Figure 8]{$\rho$, v, T, and L versus mass for the three models in figure 7 are examined at different times. Plots a) and c) show the density and temperature versus mass at time = 0 (solid line), 36 hours (dashed line), and 
47 days (dotted line). The shock can be seen at time = 36 hours in both the density and temperature profiles. The 
recombination wave can be seen at time = 47 days in the temperature profile. Plot b) shows the final velocity profile for
all three models. Plot d) shows the luminosity versus mass at time = 94 days, 190 days, and 379 days. The 
Ni heating supplies all of the energy by day 190 and by day 379 the models start to diverge showing signs of escaping gamma rays.}
\end{figure}

\begin{figure}
\plotone{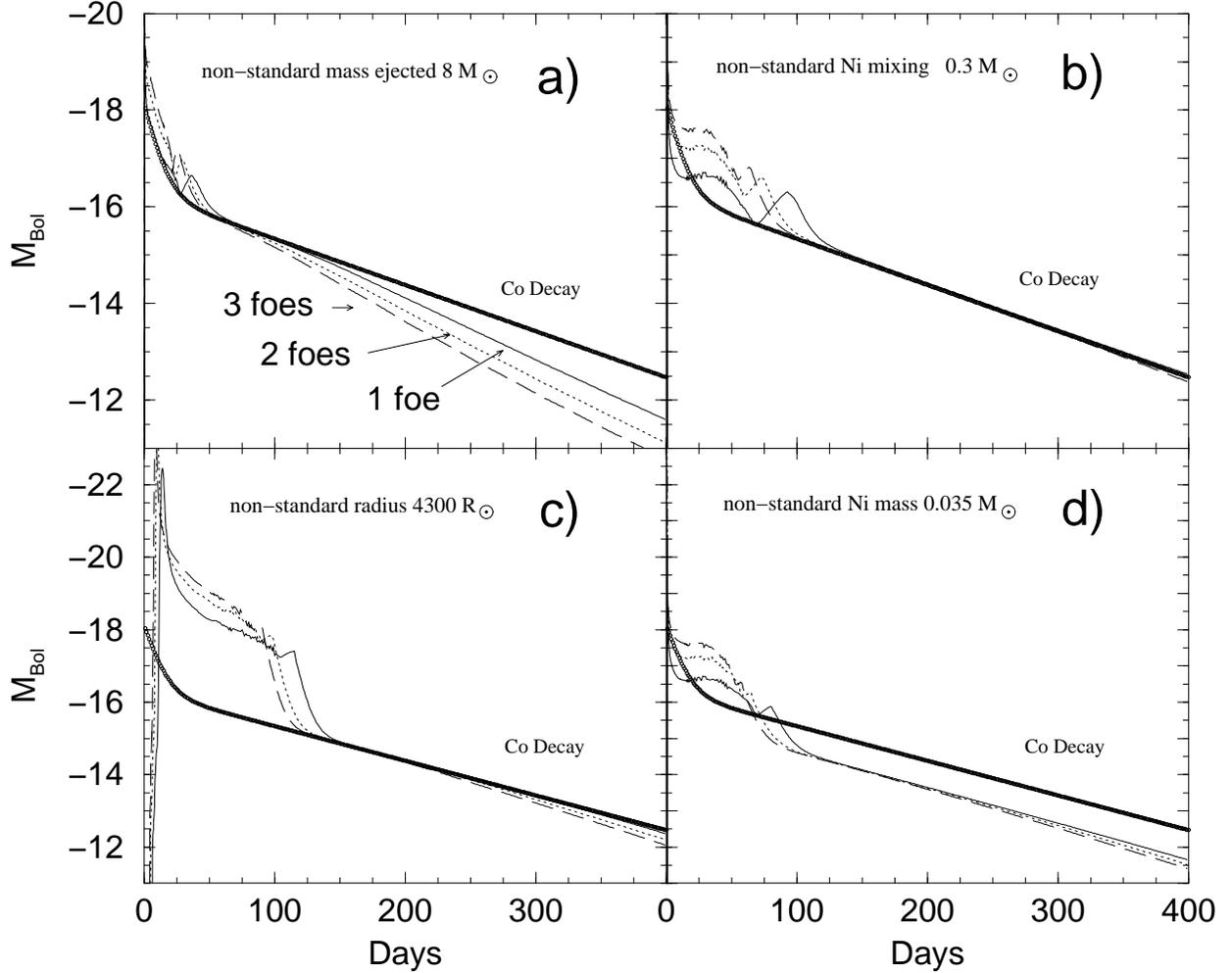}
\caption[Figure 9]{Absolute bolometric light curves showing the variations of the explosion energy with different constant values of the parameters. For 
reference the spontaneous luminosity of the Ni-CO-Fe decay for 0.07 M$_\odot$ Ni is shown. 
Plot a) shows the affect of explosion energy with the ejected mass reduced to 8 M$_\odot$ (variations of model G). Plot b) Shows the light curves of the same models as in figure 7 except the Ni mixing for all three models is $<$ .3 M$_\odot$. Plot c) shows the affect of explosion energy with the progenitor radius is increased to 4300 R$_\odot$ (variations of model C). 
Plot d) Shows the light curves of the same models as in figure 7 except the Ni mass for all three models is reduced to 0.035 M$_\odot$.
}
\end{figure}

\begin{figure}
\plotone{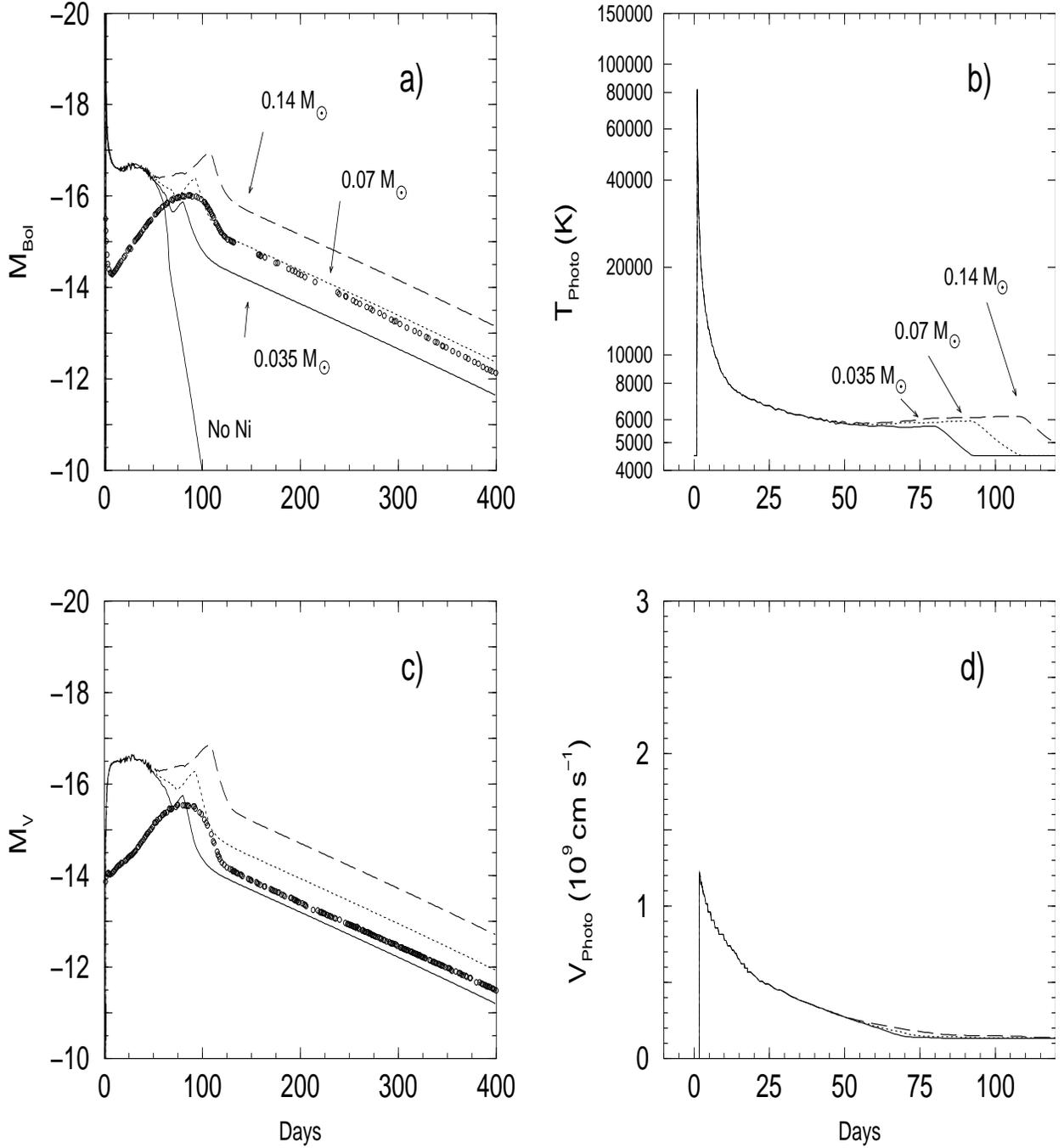}
\caption[Figure 10]{Illustrates the affect of the Ni mass on the light curve, photospheric temperature and velocity
versus time for models B with M$_{Ni}$= 0.14 M$_\odot$ (dashed line), B with M$_{Ni}$= 0.07 M$_\odot$ (dotted line), B with M$_{Ni}$= 0.035 M$_\odot$  (solid line), and Model B with no Ni mass (labeled No Ni). All models have 
the same progenitor radius, ejected mass, explosion energy, and Ni mixing: R = 430 R$_\odot$, M = 16 M$_\odot$, E = $1\times10^{51}$ ergs, Ni mixing throughout the core. The plots a) and c) show the absolute magnitude bolometric 
and visual light curves of each model
compared to SN 1987A, respectively. Plots b) and d) show the photospheric temperature and velocity of each model versus time,
respectively.}
\end{figure}

\begin{figure}
\plotone{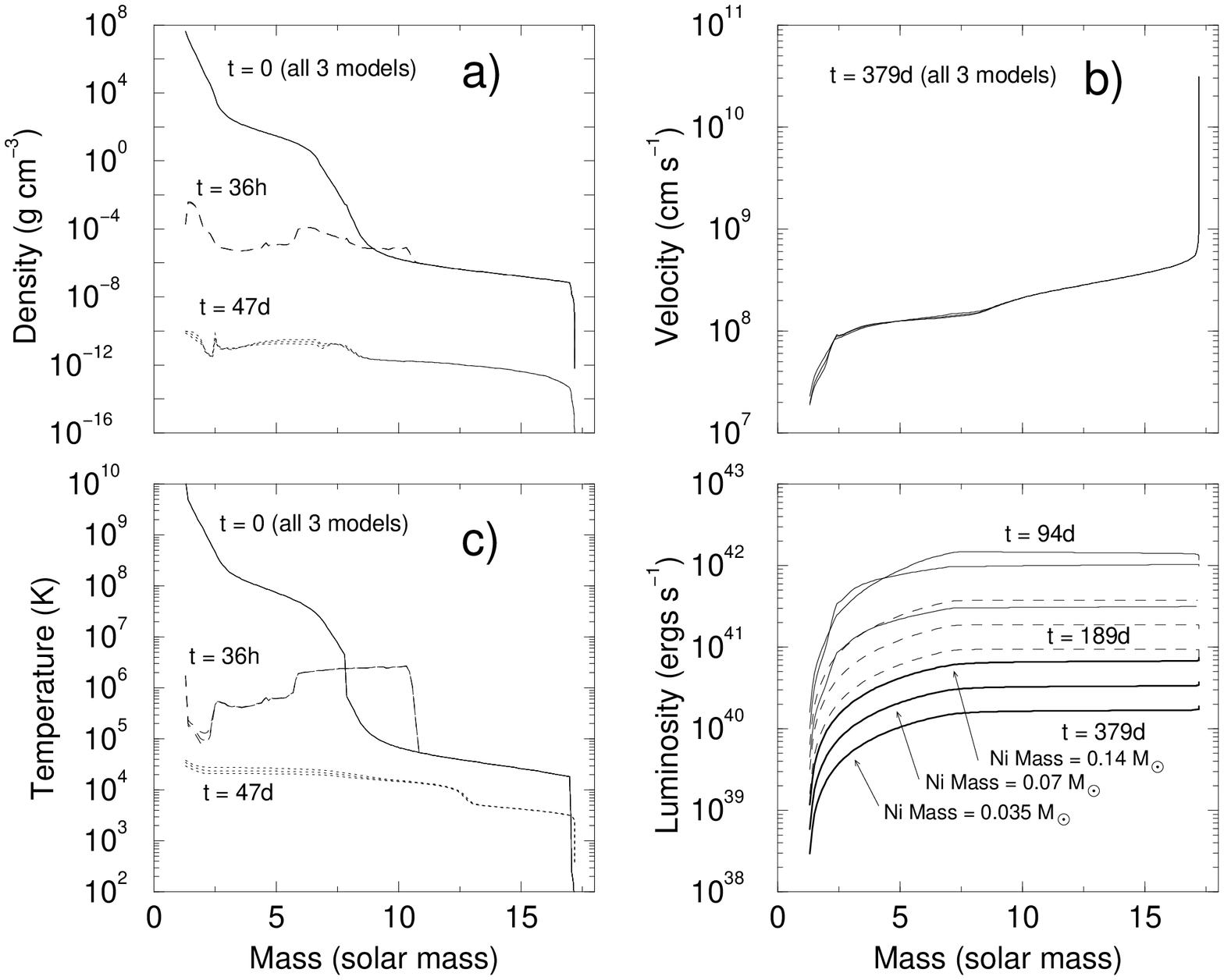}
\caption[Figure 11]{$\rho$, v, T, and L versus mass for the three models in figure 10 are examined at different times. Plots a) and c) show the density and temperature versus mass at time= 0 (solid line), 36 hours (dashed line), and 
47 days (dotted line). The shock can be seen at time = 36 hours in both the density and temperature profiles. The 
recombination wave can be seen at time = 47 days in the temperature profile. Plot b) shows the final velocity profile for
all three models. Plot d) shows the luminosity versus mass at time = 94 days, 189 days, and 379 days. The 
difference in luminosity profiles reflect the respective amounts of Ni present in the ejecta.}
\end{figure}

\begin{figure}
\plotone{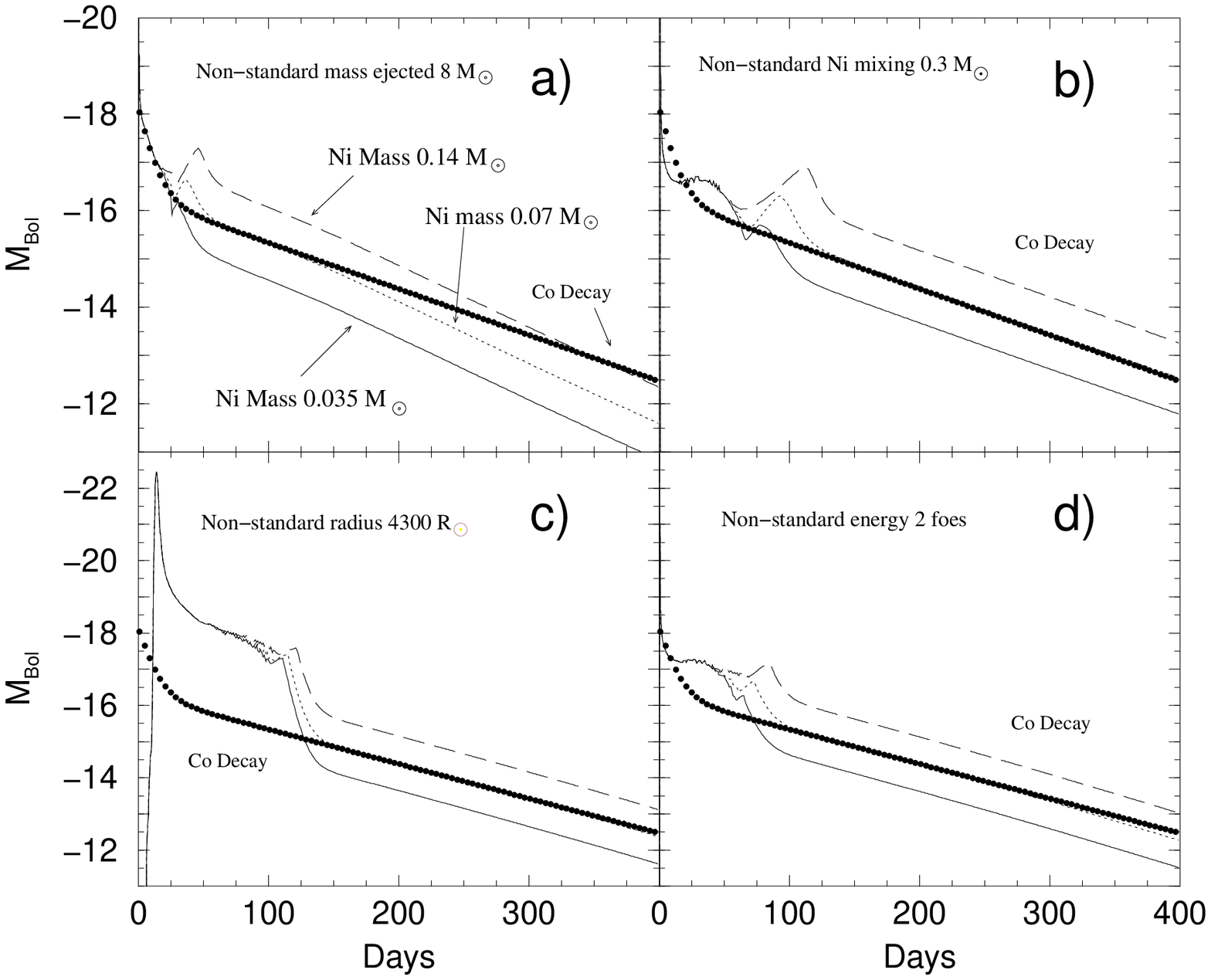}
\caption[Figure 12]{Absolute magnitude bolometric light curves showing the variations of the Ni mass with different constant values of the parameters. For 
reference the spontaneous luminosity of the Ni-CO-Fe decay for 0.07 M$_\odot$ Ni is shown. 
Plot a) shows the affect of Ni mass with the ejected mass reduced to 8 M$_\odot$ (variations of model G). 
Plot b) Shows the light curves of the same models as in figure 10 except the Ni mixing for all three models is $<$ .3 M$_\odot$. 
Plot c) shows the affect of Ni mass with the progenitor radius is increased to 4300 R$_\odot$ (variations of model C). 
Plot d) Shows the light curves of the same models as in figure 10 except the explosion energy for all three models is increased to $2\times10^{51} ergs$.}
\end{figure}

\begin{figure}
\plotone{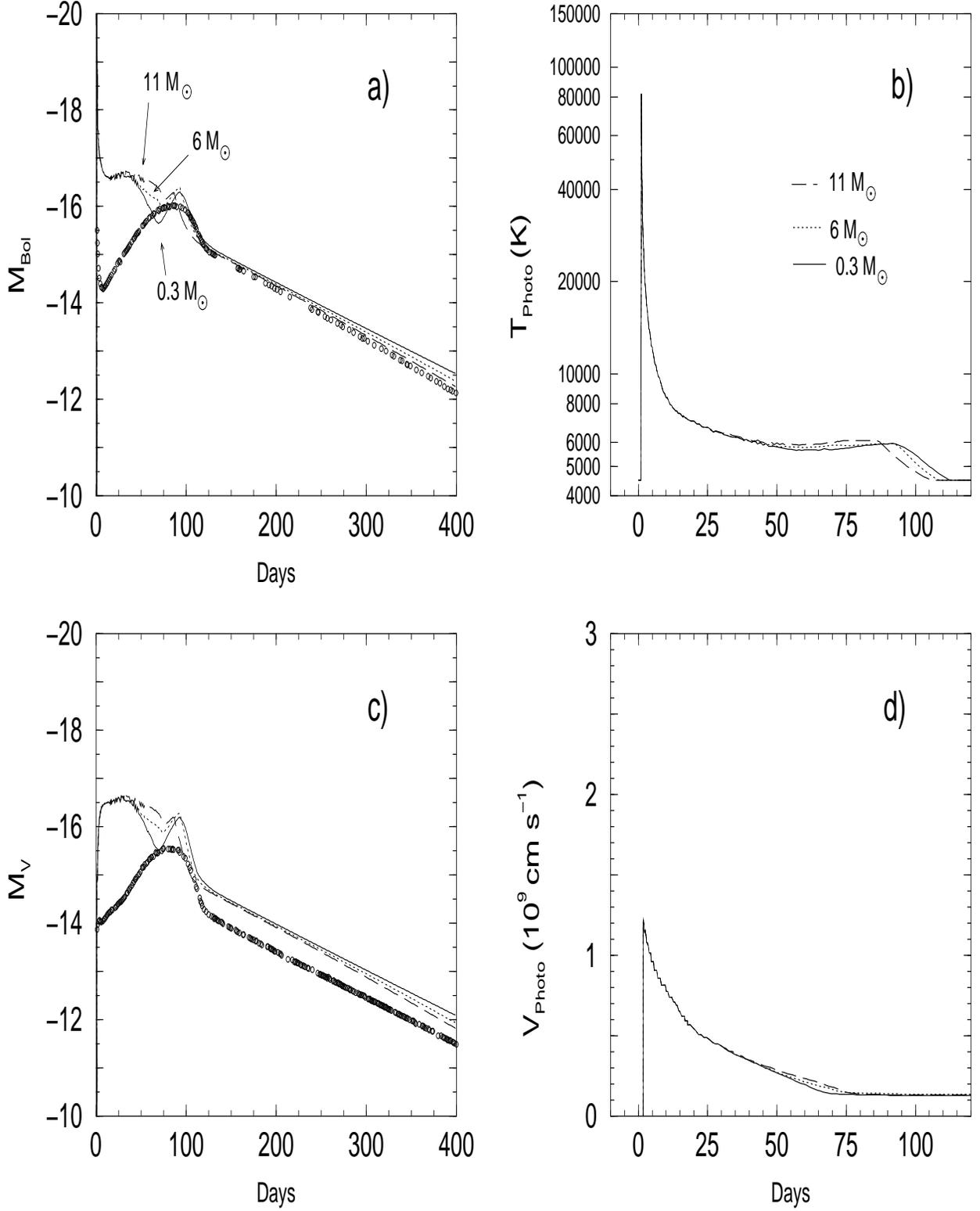}
\caption[Figure 13]{Illustrates the affect of the Ni mixing on light curve, photospheric temperature and velocity
versus time for models B with mixing to M = 11 M$_\odot$ (dashed line), B with mixing to M = 6 M$_\odot$ (dotted line), 
and B with M $<$ 0.3 M$_\odot$  (solid line). All models have 
the same progenitor radius, ejected mass, explosion energy, and Ni mass: R = 430 R$_\odot$, M = 16 M$_\odot$, E = $1\times10^{51}$ ergs, M$_{Ni}$ = 0.07 M$_\odot$. The plots a) and c) show the 
absolute magnitude bolometric and  visual light curves of each model
compared to SN 1987A, respectively. Plots b) and d) show the photospheric temperature and velocity of each model versus time
respectively.}
\end{figure}

\begin{figure}
\plotone{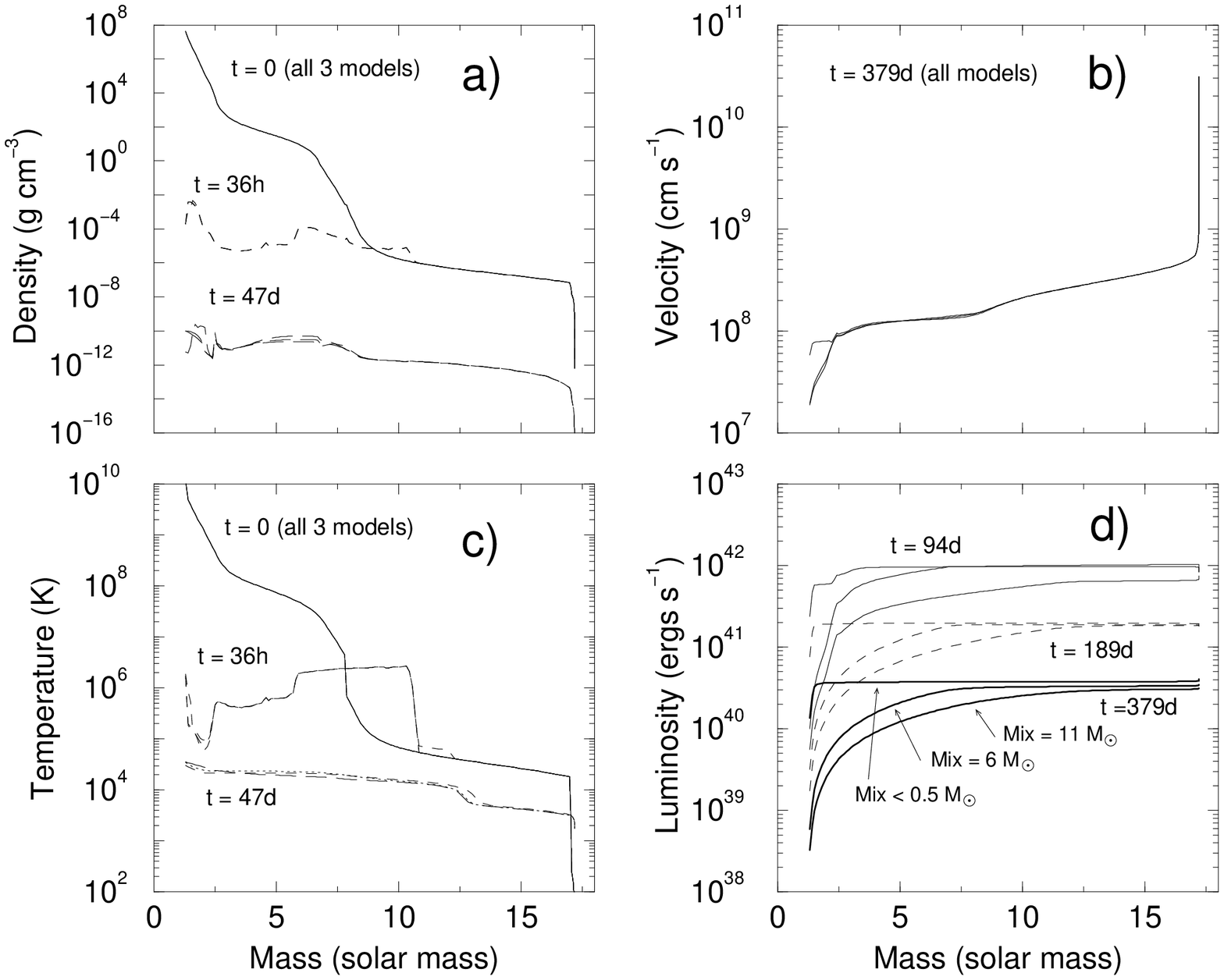}
\caption[Figure 14]{$\rho$, v, T, and L versus mass for the three models in figure 13 are examined at different times. Plots a) and c) show the density and temperature versus mass at time = 0 (solid line), 36 hours (dashed line), and 
47 days (solid line - mixing < 0.3 \smas, dotted line - mixing to 6 \smas, dashed line - mixing to 11 \smas). The shock can be seen at time = 36 hours in both the density and temperature profiles. The 
recombination wave can be seen at time = 47 days in the temperature profile. Plot b) shows the final velocity profile for
all three models. Plot d) shows the luminosity versus mass at time = 94 days, 189 days, and 379 days. The 
Ni heating supplies all of the energy by day 189 and the physical distribution of the Ni determines the 
shape of the luminosity profile,
at later times gamma rays begin to escape.}
\end{figure}

\begin{figure}
\plotone{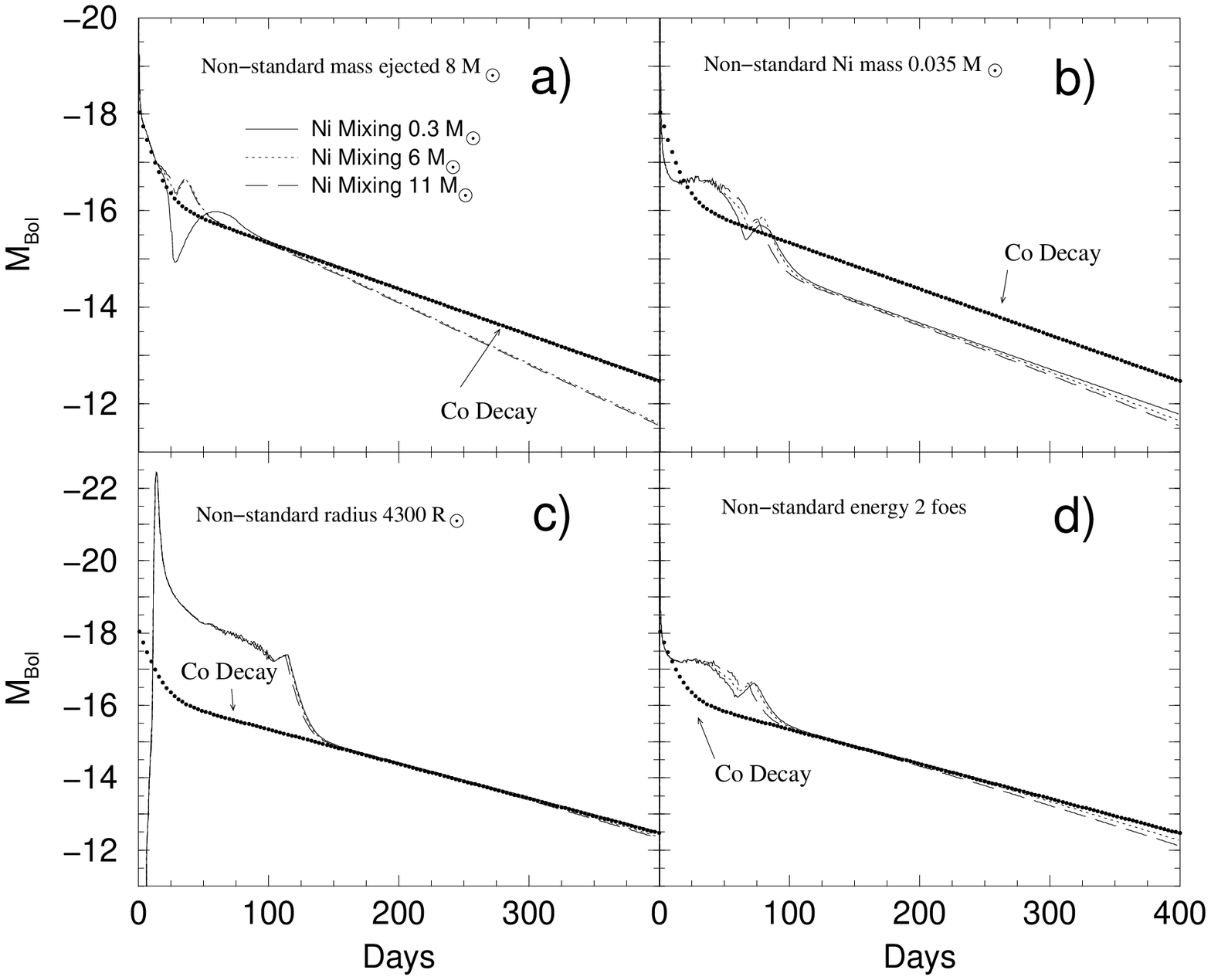}
\caption[Figure 15]{Absolute bolometric light curves showing the variations of the Ni mixing with different constant values of the parameters. For 
reference the spontaneous luminosity of the Ni-CO-Fe decay for 0.07 M$_\odot$ Ni is shown. 
Plot a) shows the affect of Ni mass with the ejected mass reduced to 8 M$_\odot$ (variations of model G). 
Plot b) shows the light curves of the same models as in figure 13 except the Ni mass for all three models is reduced to 0.035 M$_\odot$.
Plot c) shows the affect of Ni mass with the progenitor radius is increased to 4300 R$_\odot$ (variations of model C).
Plot d) Shows the light curves of the same models as in figure 13 except the explosion energy for all three models is increased to $2\times10^{51} ergs$.}
\end{figure}

\begin{table}
\caption[Model parameters]{Model Parameters}
\vskip0.05truein
\hrule
\vskip0.01truein
\hrule
\begin{tabbing}
xxxxxxxx\=xxxxxxxxxxxx\=xxxxxxxxx\=xxxxxxxxxxx\=xxxxxxxxx\=xxxxxxxx\=\kill
{\bf Model} \> {\bf Radius} \> {\bf Mass} \> {\bf Energy} \> {$\bf M_{Ni}$} \> {\bf Ni mixing} \\
{} \> {cm} \> {\smas} \> {$\times10^{51}$ ergs} \> {\smas} \>{\smas} \\
\end{tabbing}
\hrule
\begin{tabbing}
xxxxxx\=xxxxxxxxxxxxxx\=xxxxxxx\=xxxxxxxxxx\=xxxxxxxxxxxxx\=xxxxxxxxxx\=\kill
A \> $3\times10^{12}$ \> 16 \> 1,2  \> 0.035,0.07  \> center,core \\ 
B \> $3\times10^{13}$ \> 16 \> 1,2,3  \> 0.035,0.07,0.14  \> center,core,env. \\ 
C \> $3\times10^{14}$ \> 16  \> 1,2,3  \> 0.035,0.07,0.14  \> center,core,env. \\ 
D \> $3\times10^{13}$  \> 12 \> 1,2   \> 0.035,0.07,0.14 \> center,core \\ 
E \> $3\times10^{14}$  \> 12 \> 1  \> 0.07 \> core \\ 
F \> $3\times10^{12}$  \> 8  \> 1  \> 0.07  \> core \\ 
G \> $3\times10^{13}$  \> 8  \> 1,2,3  \> 0.035,0.07,0.14  \> center,core,env. \\ 
H \> $3\times10^{14}$  \> 8  \> 1 \> 0.07  \> core \\ 
\end{tabbing}
\end{table}

\end{document}